\documentclass[letterpaper]{JHEP3}
\pdfoutput=1

\usepackage[pdftex]{graphicx}
\usepackage{afterpage}
\def\simgt{\mathrel{\lower2.5pt\vbox{\lineskip=0pt\baselineskip=0pt
           \hbox{$>$}\hbox{$\sim$}}}}
\def\simlt{\mathrel{\lower2.5pt\vbox{\lineskip=0pt\baselineskip=0pt
           \hbox{$<$}\hbox{$\sim$}}}}

\title{Non-relativistic effective theory of dark matter direct detection}

\author{JiJi Fan\\
        Department of Physics, Princeton University, Princeton, NJ, 08540\\
        E-mail: \email{jijifan@princeton.edu}}
\author{Matthew Reece\\
 Princeton Center for Theoretical Science \\
 Princeton University, Princeton, NJ, 08540\\
 E-mail: \email{mreece@princeton.edu}}
 \author{Lian-Tao Wang\\
        Department of Physics, Princeton University, Princeton, NJ, 08540\\
        E-mail: \email{lianwang@princeton.edu}}
\abstract{Dark matter direct detection searches for signals coming from dark matter scattering against nuclei at a very low recoil energy scale $\sim$ 10 keV. In this paper, a simple non-relativistic effective theory is constructed to describe interactions between dark matter and nuclei without referring to any underlying high energy models. It contains the minimal set of operators that will be tested by direct detection. The effective theory approach highlights the set of distinguishable recoil spectra that could arise from different theoretical models. If dark matter is discovered in the near future in direct detection experiments, a measurement of the shape of the recoil spectrum will provide valuable information on the underlying dynamics.  We bound the coefficients of the operators in our non-relativistic effective theory by the null results of current dark matter direct detection experiments. We also discuss the mapping between the non-relativistic effective theory and field theory models or operators, including aspects of the matching of quark and gluon operators to nuclear form factors.}

\newcommand{\beq}{\begin{equation}}% can be used as {equation} or {eqnarray}
\newcommand{\eeq}{\end{equation}}
\newcommand{\beqs}{\begin{eqnarray}}% can be used as {equation} or {eqnarray}
\newcommand{\eeqs}{\end{eqnarray}}

\newcommand{\gev}{\,\mathrm{GeV}}

\newcommand{\calo}{{\cal{O}}}
\newcommand{\tih}{\tilde{h}}
\newcommand{\til}{\tilde{l}}

\begin{document}

\section{Introduction}
\label{sec: introduction}
The existence of dark matter (DM) in the Universe is well established by astrophysical and cosmological observations. However, these observations only detect DM through the universal interaction of gravity, leaving the nature of DM and its interactions almost completely unknown. Attempts to understand DM more directly by looking for signals of its interaction with the Standard Model (SM) are taking place on a variety of experimental and observational frontiers. Among them, direct detection searches for signals from DM scattering off nuclei in underground detectors. In these experiments, signals have traditionally been assumed to come from the simplest form of elastic scattering, a contact interaction independent of momentum exchange. This assumption leads to the constraints from the present null results of most direct detection experiments, usually presented as bounds on the spin independent (SI) and spin dependent (SD) cross sections per nucleon as a function of DM mass.  

The simple assumption of a DM contact interaction with nucleons overlooks the possible sensitivity of direct detection to more general DM scenarios. This has already been pointed out in the context of inelastic DM \cite{TuckerSmith:2001hy, Chang:2008gd} and form factor DM~\cite{Chang:2009yt, Feldstein:2009tr}.  In general, the form of the DM--nucleus scattering amplitude depends on properties of the dark matter particle as well as the mediator of the interaction. As we will show in this paper, other interactions could also generate sizable direct detection signals. Considering the large number of existing DM models or effective field theory operators for DM with different spins, it may appear difficult to exhaustively categorize all possible interactions that could be tested by direct detection. However, scattering in direct detection is non-relativistic (NR) as the incoming DM velocity is $v/c \sim 10^{-3}$. Thus the incident DM kinetic energy and recoil energy are around 10 keV. At this low energy scale probed by direct detection, many apparently different microscopic models or field theory operators lead to the same simple NR effective theory.  In other words, the DM--nucleus scattering in direct detection can be essentially described by an NR effective potential with small expansion parameters: the DM velocity $v \sim 10^{-3}$ and $|\vec{q}|/\Lambda$. $|\vec{q}| \sim \calo(10$ - 100 MeV) is the momentum exchange and $\Lambda$ is some large scale involved, such as the DM mass $m_\chi$, the nucleus mass $m_N$, or a heavy mediator mass. 

This NR effective theory description serves as a systematic approach to parameterize the direct detection signals for various underlying microscopic DM theories. It captures the most important differences in the measured recoil spectrum for different classes of DM models. Thus it is the simplest theory with the minimal set of operators that will be tested by the direct detection. Current direct detection results bound the coefficients of operators in the NR effective potential. Constraints on specific DM models or field theory operators could be obtained by mapping them onto the NR effective theory.

This paper is organized as follows: in section \ref{sec: effective theory}, we present the NR effective theory for direct detection and complete the various NR operators into microscopic models. In section \ref{sec: constraints}, we apply current direct detection data to constrain the Wilson coefficients of our NR theory; we also discuss constraints from other experiments on different force carriers between DM and nucleus. In section \ref{sec:nuclear}, we discuss the procedure of mapping from relativistic quark or gluon operators to NR nucleus operators. We conclude in section \ref{sec: conclusion}. Finally, we include formulae for the nuclear recoil rate, power counting rules and the mapping between NR theory and field theory operators in appendices.

\section{NR effective theory for direct detection}
\label{sec: effective theory}
\subsection{General considerations}
\label{sec: general}

The DM velocity in the galactic halo is typically of the order $v \sim 10^{-3}$. For DM and target nuclei with mass around the weak scale, this implies incident DM kinetic energy and recoil energy around 10 keV, much smaller than typical nuclear binding energy (1 - 10) MeV per nucleon. This justifies the treatment of DM scattering against the whole nucleus using NR quantum mechanics. In our framework, the DM-nucleus interaction is described by an effective potential, $V_{\rm eff}$. It is a rotationally invariant scalar formed out of four 3-vectors: the relative velocity $\vec{v}$, DM position $\vec{r}$, DM spin $\vec{s}_{\chi}$ and nuclear spin $\vec{s}_N$. The case in which $\vec{s}_N$ does not appear is known as spin-independent (SI) scattering, as opposed to spin-dependent (SD) scattering. To the first order of the Born approximation, the amplitude in the NR limit is the Fourier transform of the effective potential in position space:
\beq
{\cal{M}}(\vec{q},\vec{v}) = - \int d^3\vec{r} e^{i\vec{q}\cdot\vec{r}}V_{\rm eff}(\vec{r},\vec{v}),
\label{eq: matrix element}
\eeq
where $\vec{q}$ is the transferred momentum. In general, the $\vec{r}$ dependence will involve the structure of the nucleus, due to its finite size. This effect requires a nuclear form factor depending on the transferred momentum $\vec{q}$ in the calculation of the differential recoil rate. In tabulating the operators that can appear in $V_{\rm eff}$, we will factor out $A F(q)$, the number of nucleons times the nuclear form factor, in the SI case, and $\sqrt{S(q)/S(0)}$, the spin form factor, in the SD case. (So, for instance, if the DM interacts with the charge of the nucleus, we factor out $A F(q)$ and the coupling will be rescaled by a factor of $Z/A$.) We will discuss more details of the role of nuclear physics and the form factors that can appear in Sec.~\ref{sec:nuclear}, and give explicit examples of the matching procedure in detail in App.~\ref{sec:example} to clarify our conventions.

The mass scales that enter into the potential are the DM mass $m_{\chi}$, the nucleus mass $m_{N}$, and the mediator mass $m_{\phi}$. In addition, there could be other scales $\Lambda^\prime$ present. For instance, the DM-mediator interaction could arise at the nonrenormalizable level, encoded by a high dimensional operator suppressed by powers of $\Lambda^\prime$. This could happen, for example, in models of DM with zero electric charge but higher-order electromagnetic form factors. Notice that direct detection experiments constrain one combination of the mass scales and the coupling constants. Thus, in the following parameterization of NR effective theory, we choose to absorb all of the scale dependence into the coefficients of the terms in the effective potential.

We will consider two qualitatively different cases: a contact interaction when $|\vec{q}| \ll m_{\phi}$ and the mediator is integrated out, and a long range interaction when $|\vec{q}| \gg m_{\phi}$. The leading term in the effective potential (after factoring out the nuclear form factor) is then proportional to $\delta^3(\vec{r})$ and $1/r$, respectively. For a contact-type interaction, the NR expansion of the potential is a derivative expansion, while for a long-range interaction, it is a multipole expansion. It is often assumed that, due to the smallness of the expansion parameters, direct detection experiments will only be sensitive to a momentum-independent potential. However, as will be shown in Sec.~\ref{sec: models}, it is possible that the leading contributions from microscopic models are already suppressed by a single $|\vec{q}|$ or $|\vec{v}|$. High-order terms in $|\vec{q}|$ could still be detectable if they are leading, especially for the case of long range interactions. However, given the parameter space probed by experiments, the potentials relevant for direct detection form a finite set. In the studies below, for simplicity, we will focus on potential terms suppressed by at most a single $|\vec{q}|$. More details of power counting rules for the two cases may be found in Appendix~\ref{sec: powercounting}.

\subsection{NR effective potential}
\label{sec:NReffpot}

We begin by writing down the effective potentials $V_{\rm eff}^{\rm SI}$ for SI scattering and $V_{\rm eff}^{\rm SD}$ for SD scattering, 
\beqs
V_{\rm eff} &=& V^{\rm SI}_{\rm  eff} + V^{\rm SD}_{\rm eff} \nonumber \\
V^{\rm SI}_{\rm eff} &=& h_1\delta^3(\vec{r}) - h_2 \vec{s}_\chi\cdot \vec{\nabla}\delta^3(\vec{r}) \nonumber \\
&+&l_1{1\over 4\pi r }  +l_2{ \vec{s}_\chi\cdot \vec{r}\over 4\pi  r^3}~,  
\label{eq: Hamiltonian}\\
V^{\rm SD}_{\rm eff} &=&h_1^\prime\vec{s}_\chi\cdot \vec{s}_N\delta^3(\vec{r}) - h_2^\prime \vec{s}_N\cdot \vec{\nabla}\delta^3(\vec{r})\nonumber \\
&+&l_1^\prime{\vec{s}_\chi\cdot \vec{s}_N\over 4\pi r} +l_2^\prime{\vec{s}_N\cdot \vec{r}\over 4\pi r^3}~, 
\label{eq: Hamiltonian sd}
\eeqs
where the (dimensionful) Wilson coefficients $h$ (for heavy mediators) and $l$ (for light mediators) are determined by matching the field theory operators from the underlying theory to the NR operators. They are proportional to the couplings of DM to the mediator, $g_\chi$, as well as the couplings of the nucleus to the mediator (which in general can involve nuclear physics quantities like those traditionally denoted $f_{n,p}$ in the DM literature). One should understand the terms in the effective potential to carry spin indices, which we have suppressed; for instance, $h_1 \delta^3(\vec{r})$ multiplies $\delta_{rr'} \delta_{ss'}$, where $r,r'$ and $s,s'$ are the spins of the nucleus and DM before and after scattering. Similarly, $\vec{s}_\chi$ is to be understood as an appropriate representation of spin, $\vec{s}_{\chi;ss'}$; e.g. for fermionic DM, it will be given by the Pauli matrices in a basis with given spins, $\frac{1}{2} \xi^{\dagger(s')}\vec{\sigma}\xi^{(s)}$. Numerical factors in the translation of $V_{\rm eff}$ to a spin-averaged cross section $d\sigma/dE_R$ will depend on the spin structure. We will adopt the convention that all cross sections and limit plots are for the case that DM is a Dirac fermion; the results can be easily rescaled to other cases. Detailed examples of matching are given in section \ref{sec:example}. Note that in general, the couplings $h_1^\prime, \ldots$ will be isotope-dependent. DM models that give rise to the above NR operators have already appeared in the literature (Recent examples include momentum dependent DM~\cite{Chang:2009yt, Feldstein:2009tr} and DM with electromagnetic form factor~\cite{Pospelov:2000bq, Masso:2009mu, Sigurdson:2004zp, Kribs:2009fy, An:2010kc, Chang:2010en, Barger:2010gv, Fitzpatrick:2010br, Banks:2010eh}). The NR effective theory will manifest the experimentally testable physics without referring to diverse high energy interpretations.

The first operators in $V^{\rm SI}_{\rm eff}$ and $V^{\rm SD}_{\rm eff}$ are the most-studied cases, SI and SD contact interactions. The momentum suppressed operators are usually neglected. Yet in the absence of momentum-independent operators or with large enough coefficients to compensate the momentum suppression, these terms could still be relevant for direct detection. A couple of comments are in order:
\begin{itemize}
\item Any specific DM model typically gives a subset of the operators presented in Eq.~(\ref{eq: Hamiltonian}) and Eq.~(\ref{eq: Hamiltonian sd}). The most stringent bound from direct detection is always only on the leading order operator.  As we will show, each operator presented is the leading contribution from  a natural UV completion.   
\item We have assumed a derivative expansion in $\vec{r}$, corresponding to powers of ${\vec q}$ in momentum space. Thus, for instance, a $1/r^3$ term is omitted because it corresponds to a $\log |\vec{q}|$ interaction. Logarithmic dependence on $|\vec{q}|$ signals the exchange of a continuum of modes, and could arise if both DM and the SM couple to a new massless or conformal sector (which could have a mass scale far below the other scales in the problem). Such sectors have received attention in the guises of RS2 \cite{RS2} and unparticles \cite{Unparticles}, but we are not aware of models of such a sector mediating interactions between DM and the SM. A $\log|\vec{q}|$ dependence would have fairly mild effects on the shape of the recoil energy distribution, which is subject to uncertainties including nuclear physics and the DM velocity distribution, so we expect that it would be extremely challenging to deduce evidence for a new conformal sector solely from DM direct detection.
\item{We consider only static potentials. There are also operators depending on the relative velocity $v$ of the DM and nucleus. For instance, the operators suppressed by a single $v$ are
\beq
  V =\vec{s}_\chi\cdot \vec{v}\,\delta^3(\vec{r}),\quad \vec{s}_N\cdot \vec{v}\,\delta^3(\vec{r}), \quad { \vec{s}_\chi\cdot \vec{v}\over 4\pi r }, \quad {\vec{s}_N\cdot \vec{v}\over 4\pi r}, 
\eeq
It turns out that for elastic scattering, numerically the shapes of the recoil spectra they generate are almost identical to those of the unsuppressed operators. (Their effect is most important on the tail of the $E_R$ distribution, which is relatively unimportant.) Thus we neglect them for the rest of the studies. Experimentally, this higher-dimensional space of potentials leads to a ``degeneracy" in the inverse problem of mapping measured spectra to NR operators. We discuss this in somewhat more detail in Appendix~\ref{sec:example}.
 }
 \item{ In eqs.~(\ref{eq: Hamiltonian}, \ref{eq: Hamiltonian sd}), we only keep operators suppressed by a single $|\vec{q}|$. Our potential is the minimal extension of the well-studied momentum independent contact interaction. If the leading operator from a field theory is suppressed by more powers of  $|\vec{q}|$, it could still contribute an observable rate to direct detection, depending on the mediator mass. We will give a more exhaustive list in Appendices~\ref{sec: fermion}, \ref{sec: scalar}, \ref{sec: vector} but focus on eqs.~(\ref{eq: Hamiltonian}, \ref{eq: Hamiltonian sd}) in the numerical studies.}
\end{itemize}

\begin{table}[h]
\begin{center}
\begin{tabular}{|c|c|c|}
\hline 
SI NR operators & SD NR operators & $E_R$\\
\hline
\hline
$\delta^3(\vec{r})$ & $\vec{s}_\chi\cdot \vec{s}_N\delta^3(\vec{r})$ & $1$ \\  
$\vec{s}_\chi\cdot \vec{\nabla}\delta^3(\vec{r})$&$\vec{s}_N\cdot \vec{\nabla}\delta^3(\vec{r})$&$E_R$\\
${1\over 4\pi r }$&${\vec{s}_\chi\cdot \vec{s}_N\over 4\pi r}$ & $E_R^{-2}$ \\
${\vec{s}_\chi\cdot \vec{r}\over 4\pi r^3}$ &${\vec{s}_N\cdot \vec{r}\over 4\pi r^3}$ &$E_R^{-1}$ \\
\hline
\end{tabular}
\caption{Recoil energy dependence of effective cross section per nucleon for the NR operators. (Nuclear form factors are factored out.)}
\label{table:Er dep}
\end{center}
\end{table}

As emphasized in the introduction, the effective theory parameterization highlights the possibility of having qualitatively different recoil energy spectra, as shown in Table~\ref{table:Er dep}. 

\subsection{Connection with microscopic models}
\label{sec: models}

The matching between microscopic models and effective potentials goes through three straightforward steps. Starting with a particular DM model, we first write down the relativistic field theory operators relevant for the scattering process, $G(q^2,m_{\phi}) J_{\chi}  J_{q(g)}$. $J_{\chi}$ and $J_{q(g)}$ are appropriate DM and quark (gluon) operators, respectively. $G(q^2, m_{\phi})$ comes from the exchange of the mediator.  Second, we convert this to an operator involving nucleons by taking the nuclear matrix element $\left<N(p+q)|J_{q(g)}(q)|N(p)\right>$. Then we take the Fourier transform of the scattering amplitude in the NR limit to get the effective potential, factoring out the nuclear form factor in the definition of $V_{\rm eff}$.  We present the result of the final step matching in the cases of scalar, fermion and vector DM in Appendix~\ref{sec: fermion}, \ref{sec: scalar}, \ref{sec: vector}. Details of the nuclear matrix elements are discussed in section \ref{sec:nuclear}. In this section, we discuss simple high energy models of fermion DM for the set of NR operators in Eq.~(\ref{eq: Hamiltonian}, \ref{eq: Hamiltonian sd}).

 Most of the existing DM models yield simple contact interactions as the leading operator. It is possible, however, the coefficients of momentum (velocity) suppressed operators dominate over that of the simplest contact interaction. The coefficient enhancement could be due to large couplings of a mediator to the dark sector or small mediator mass as $\sigma \sim m^{-4}$. In models where the leading operator is SD while the SI operators are momentum-suppressed, the SI scattering could still be detectable as SI searches probe much weaker processes than SD searches. For the SI operators, we have

\begin{itemize}
\item{$\mathbf{\delta^3(\vec{r})}$ \\
This is the most studied case. For instance, Higgs exchange between fermion DM and the nucleus would lead to an operator $\bar{\chi}\chi\bar{q}q$. With Higgs mass around 100 GeV and the Higgs--nucleon coupling about the strange-quark Yukawa coupling $10^{-3}$, this would lead to a plausible scattering cross section $\sim 10^{-44}$ cm$^2$.  This could also arise from gauge boson exchange, e.g., a $Z$ boson exchange. The DM--$Z$ coupling has to be of order $10^{-3}$ for a $10^{-44}$ cm$^2$ cross section.  This small coupling could come from some high dimensional operator. For instance, assume that DM and the SM higgs are both charged under a new $U(1)$ which is broken at  scale $\Lambda \sim$10 TeV. Integrating out the $Z^\prime$ leads to a dimension six operator $\bar{\chi}\gamma^\mu \chi h^\dagger D_\mu h$ which induces an effective DM-$Z$ coupling of order $(v_{EW}/\Lambda)^2 \sim (10^{-4} - 10^{-3})$. }

\item{$\mathbf{\vec{s}_\chi\cdot\vec{\nabla} \delta^3(\vec{r})}$\\
This operator could arise from $d^\prime \bar{\chi} \sigma^{\mu\nu} \gamma^5 \chi F_{\mu\nu}^\prime$, an electric dipole coupling of DM to a new GeV gauge boson which kinetically mixes with the photon $\epsilon F_{\mu\nu}^\prime F^{\mu\nu}$ \cite{Holdom}. The NR operator coefficient $h_2$ is related to the dark electric dipole moment $d^\prime$ as $h_2 \sim d^\prime \epsilon / m_\phi^2$. The direct detection bound $h_2 \lesssim 10^{-7}$ GeV$^{-3}$ becomes $d^\prime \epsilon \lesssim 10^{-21}$ (e cm). Notice that this interaction leads to a potential proportional to the charge $Z$ of the nucleus instead of the atomic number $A$. 

This operator could also be generated by $i\bar{\chi}\gamma^5\chi\bar{q}q$ from heavy scalar/pseudoscalar exchange. In this case, $h_2 \sim g/(m_\chi m_\phi^2)$, where the dimensionless coefficient $g$ is proportional to the CP-odd coupling from either the DM sector or the visible sector. For DM with mass around the weak scale $m_\chi \sim \calo(100$ GeV), $g \sim \calo (0.1)$ for a 100 GeV mediator and $g \sim \calo(10^{-5})$ for a 1 GeV mediator.

This operator could also appear in a special linear combination with the velocity-suppressed operator $\vec{s}_\chi \cdot \vec{v}$, giving a shape that is similar to a linear combination of the shapes from couplings $h_1$ and $h_2$. The example is the DM ``anapole" moment scattering off the nuclear electric current, which will be discussed in detail in Appendix~\ref{sec:example}.}

\item{$\mathbf{1/r}$ \\
The Coulomb potential could arise through exchange of some new light bosons with $m_\phi < |q|$. The current direct detection bounds require the coupling to be tiny, $\simlt 10^{-11}$ (see Sec.~\ref{sec: other constraints}). Although such a coupling may seem unreasonably small, it is not difficult to build dark photon models that satisfy the bound, e.g. the new light gauge boson kinetically mixed with the photon only through $S$-parameter-like higher dimension operators, as discussed in \cite{ArkaniHamed:2008qp}. }

\item{$\mathbf{\vec{s}_\chi\cdot\vec{r}/r^3}$ \\
This is the DM-dipole/nucleus-monopole coupling. They could arise from models similar to those that generate the contact interaction $\vec{s} \cdot \vec{\nabla} \delta^3(\vec{r})$, but with the mediators lighter than the momentum transfer. For instance, DM electric dipole coupling to the ordinary photon would give this potential with coefficient $l_2 \sim d$. The direct detection bound $l_2 \lesssim 10^{-9}$ GeV$^{-1}$ gives the DM electric dipole moment $d \lesssim 10^{-23}$ (e cm). For models with light scalar or pseudoscalar exchange, $l_2 \sim g/m_\chi$ leading to the CP odd coupling $g \lesssim 10^{-7}$ for weak scale DM.

Now we would like to estimate roughly how large the CP violating coupling could be if the CP violation is confined to the visible sector. CP-violating phases in the visible sector could be present in various extensions of the SM. They generically have to be small to avoid generating electric dipole moments for the neutron, electron, and atoms in conflict with observed data. For instance, in the minimal supersymmetric SM, a combination of complex phases of the gaugino-mass parameters, the A parameters, and $\mu$ must be less than the order of $10^{-2} -10^{-3}$ (for a supersymmetry-breaking scale of 100 GeV). If a CP violating coupling like $\pi^\prime \bar{q}q$ with $\pi^\prime$ a pseudoscalar is generated at one loop, the coupling may be further suppressed by the loop factor. Taking into account other possible small couplings, one could get values in the range $10^{-5} -10^{-7}$. Intriguingly, this is around the bound set by direct detection. }

 \end{itemize}
 
  For the SD operators, the leading contact interaction $\vec{s}_\chi \cdot \vec{s}_N \delta^3(\vec{r})$ again arises from most existing models, e.g, from $Z$ exchange or squark exchange in the neutralino DM scenario. If the mediator gauge boson is light, it may lead to the long range SD potential $\vec{s}_\chi \cdot \vec{s}_N/r$. Consider the case that the mediator is a pseudoscalar with CP-violating coupling to the dark sector $\pi^\prime \bar{\chi} \chi$ due to some hidden CP violating phases. If the mediator is heavy, it will lead to $\vec{s}_N\cdot\vec{\nabla} \delta^3(\vec{r})$, while if it is light, it generates the nucleus-dipole/DM-monopole potential $\vec{s}_N \cdot \vec{r}/r^3$.

\section{Constraints and sensitivities}
\label{sec: constraints}

\subsection{Constraints from current direct detection experiments}
\label{sec: numerics}
In this section we constrain the coefficients appearing in Eq.~(\ref{eq: Hamiltonian}, \ref{eq: Hamiltonian sd}) using the most sensitive direct detection experiments. We consider two classes of experiments based on different techniques. One class of experiments is based on measurement of the energy DM deposits in a detector by scattering off target nuclei. Most direct detection experiments, such as CDMS-II \cite{Ahmed:2008eu,Ahmed:2009zw}, XENON10 \cite{Angle:2007uj, Angle:2009xb, Angle:2008we}, and XENON100 \cite{Aprile:2010um}, belong to this class. Another class of experiments, especially useful for the SD constraints, is based on the superheated droplet (bubble chamber) technique to search for DM recoiling on $^{19}F$ nuclei in a compound target. This process is very sensitive to SD interactions. If the energy deposited by DM exceeds a minimal energy barrier determined by the experimental thermodynamical conditions, it will trigger the nucleation of a bubble of the gas phase in the superheated liquid. The experiments search for bubble formation events not accounted for by various backgrounds. They constrain the total rate of DM--nucleus scattering but cannot give the differential rate as the first class of experiments do. COUPP \cite{coupp}, SIMPLE \cite{Felizardo:2010mi} and PICASSO \cite{Archambault:2009sm} are the leading experiments in this class. 

For the SI constraints, we use data from CDMS-II \cite{Ahmed:2008eu, Ahmed:2009zw}, XENON10 \cite{ Angle:2007uj, Angle:2009xb}, and the currently available results of XENON100 \cite{Aprile:2010um}. For the SD case, we use data from XENON10 and XENON100 for the DM--neutron coupling, as half of naturally occurring xenon is in the form of isotopes with unpaired neutrons \cite{Angle:2008we}. We use the preliminary results of COUPP \cite{coupp} to constrain the SD DM--proton coupling. For CDMS-II, XENON10, and XENON100, we will use the maximum gap method \cite{Yellin:2002xd} to set limits while for COUPP, we will take the 90\% CL Poisson limit from their three DM candidate events, allowing $\sim$ 6 events. In our calculations we have used form factors from Refs. \cite{Duda:2006uk,Bednyakov:2006ux}.

We use the parametrization of Eq.~(\ref{eq: Hamiltonian}, \ref{eq: Hamiltonian sd}) for numerical studies. The formulae for recoil rates are presented in Appendix~\ref{app: direct detection}. We set one coefficient to be nonzero at a time and plot the spectrum in Fig.~\ref{fig: spectrum}. From the left panel of Fig.~\ref{fig: spectrum}, one can easily see that the cross section scaling as $E_R$ peaks toward high energy and is broader than the other distributions. The other rates peak at the threshold and have different slopes determined by their $E_R$ power dependence. In Fig.~\ref{fig: spectrum}, we also show a spectrum with two comparable contributions from different operators. The spectrum resembles that of the ``semielastic DM" scenario \cite{Krohn:2010ad}, as it peaks at the threshold but also rises at higher energies. The only difference is that in ``semielastic DM" the spectrum comes from a combination of elastic and inelastic scattering, while here we only consider elastic scattering. One could play further and consider both elastic and inelastic scattering and various combinations of operators. If, in the near future, direct detection confirms DM signals and collects enough data, one in principle could fit the recoil spectrum to our NR theory parameterization to obtain information about the DM interaction. 

\FIGURE[h]{
\begin{tabular}{ccc}
 \includegraphics[scale=0.7]{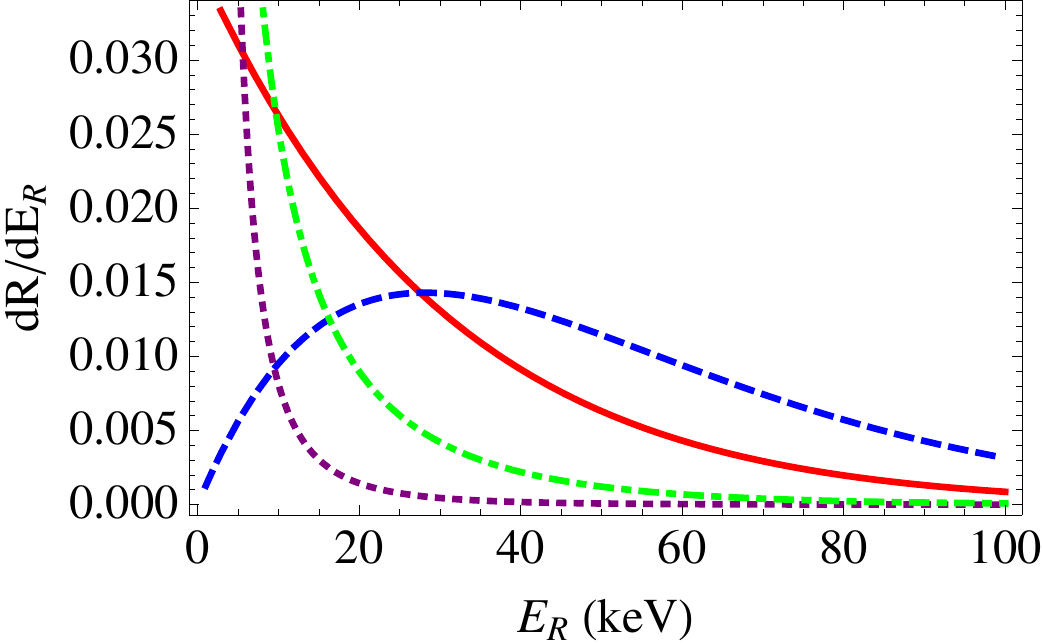}
 \includegraphics[scale=0.7]{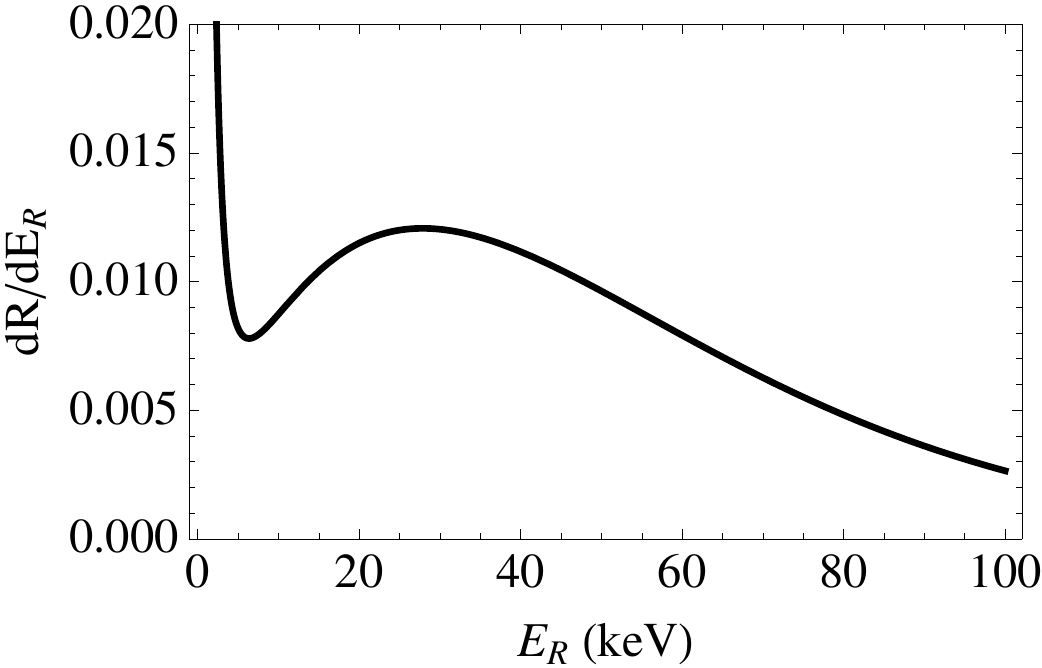}
 \end{tabular}
 \caption{Left: Recoil energy spectrum for 150 GeV DM scattering on germanium. The curves correspond to spectra with one of $h_1, h_2, l_1, l_2$ being nonzero in red, blue dashed, purple dotted, and green dot-dashed.  Right:  Recoil energy of 150 GeV DM on germanium with nonzero $h_2$ and $l_1$.  All the curves are normalized to have the same number of total events above 1 keV.}
\label{fig: spectrum} 
}

\FIGURE[h]{
\begin{tabular}{ccc}
 \includegraphics[scale=0.5]{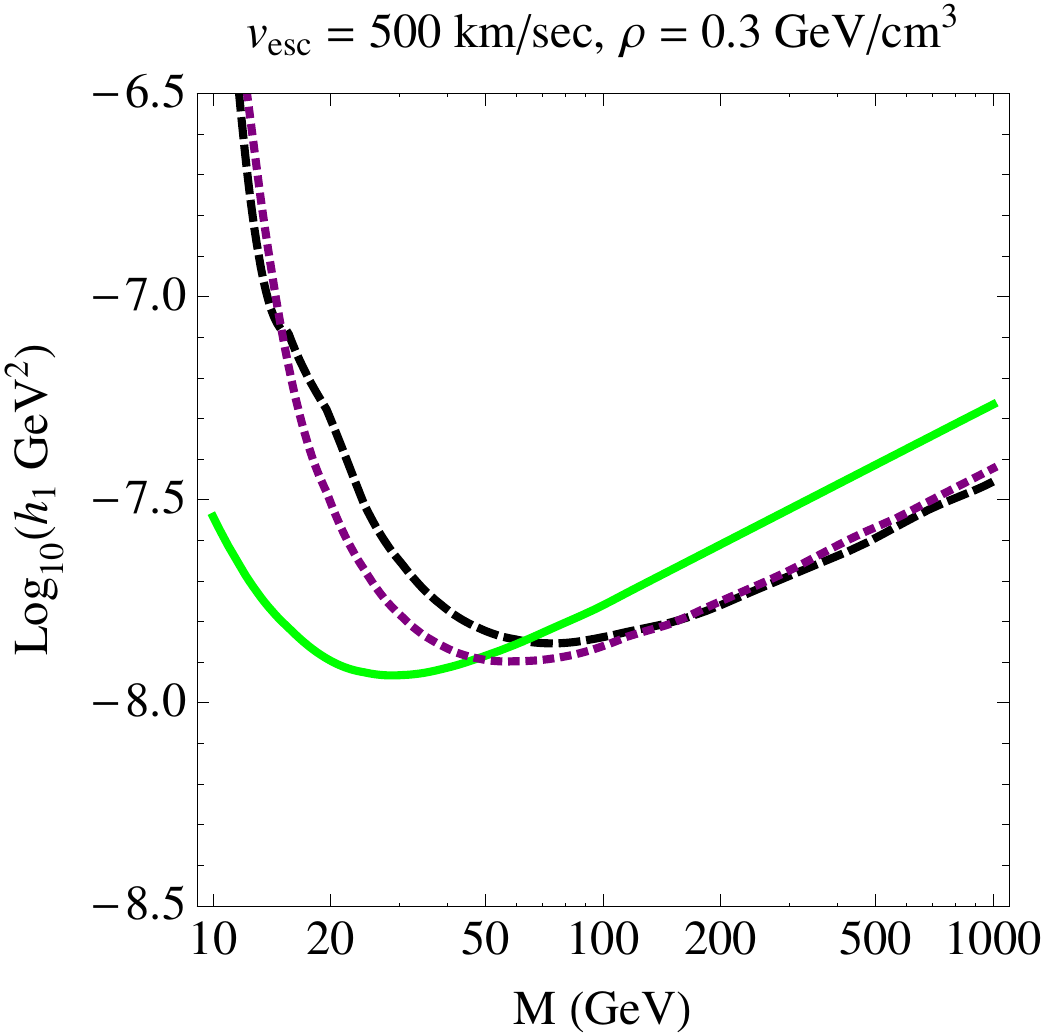}&
\includegraphics[scale=0.5]{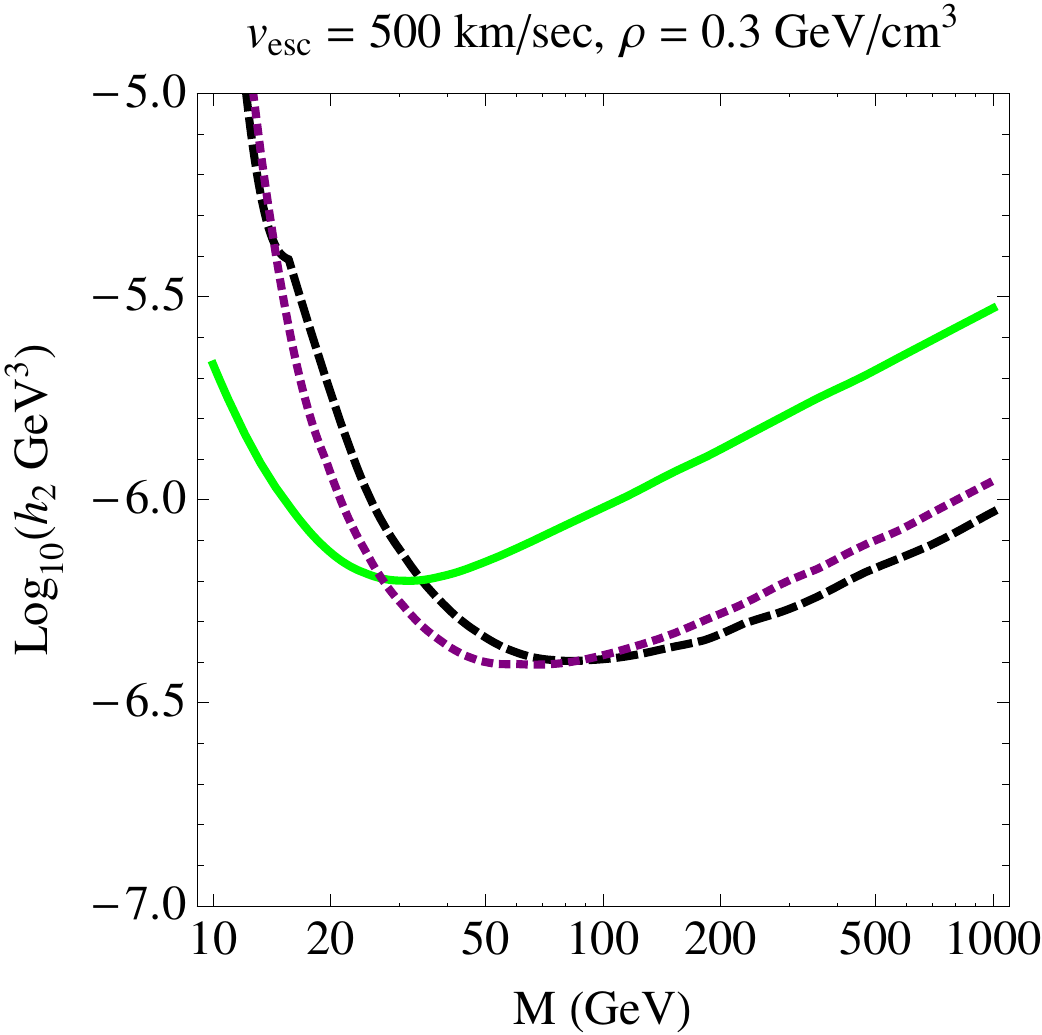}&\\
\includegraphics[scale=0.51]{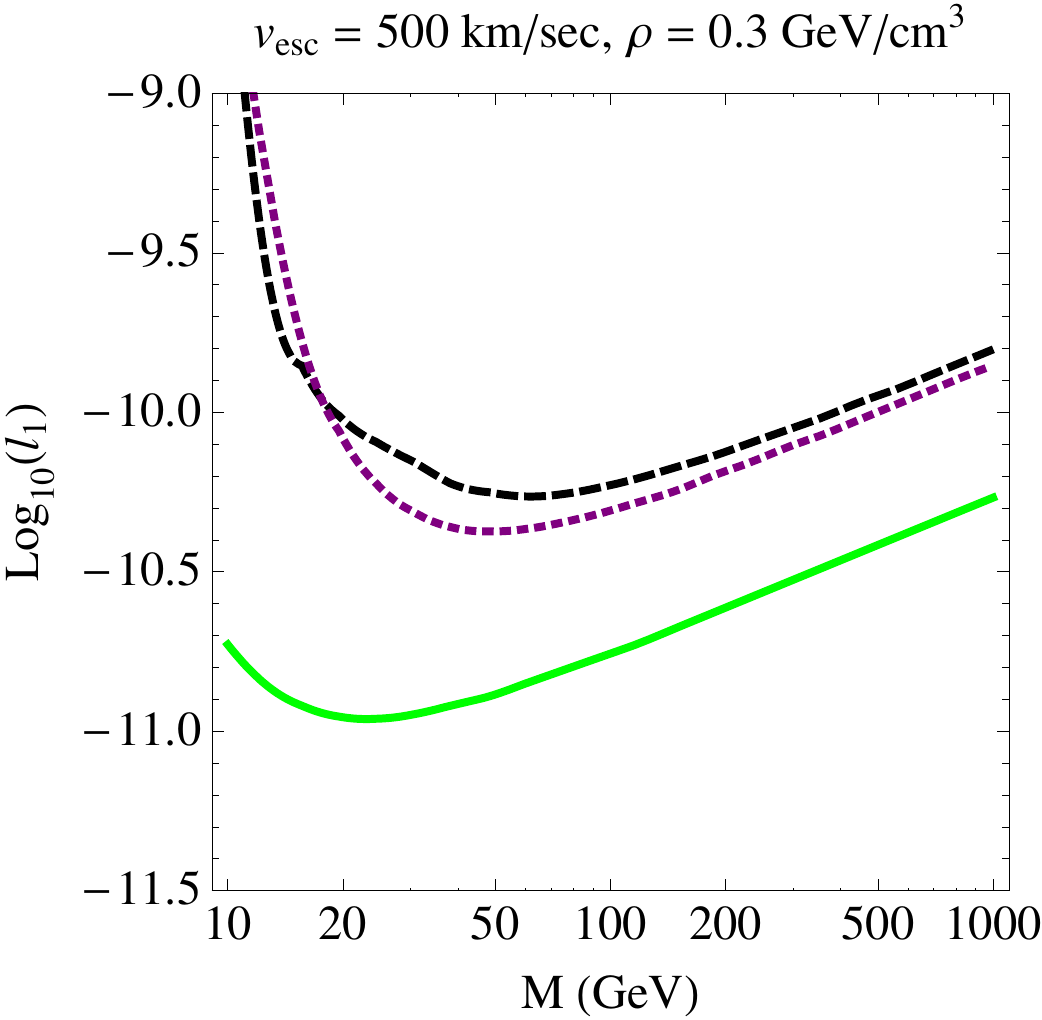}&
\includegraphics[scale=0.5]{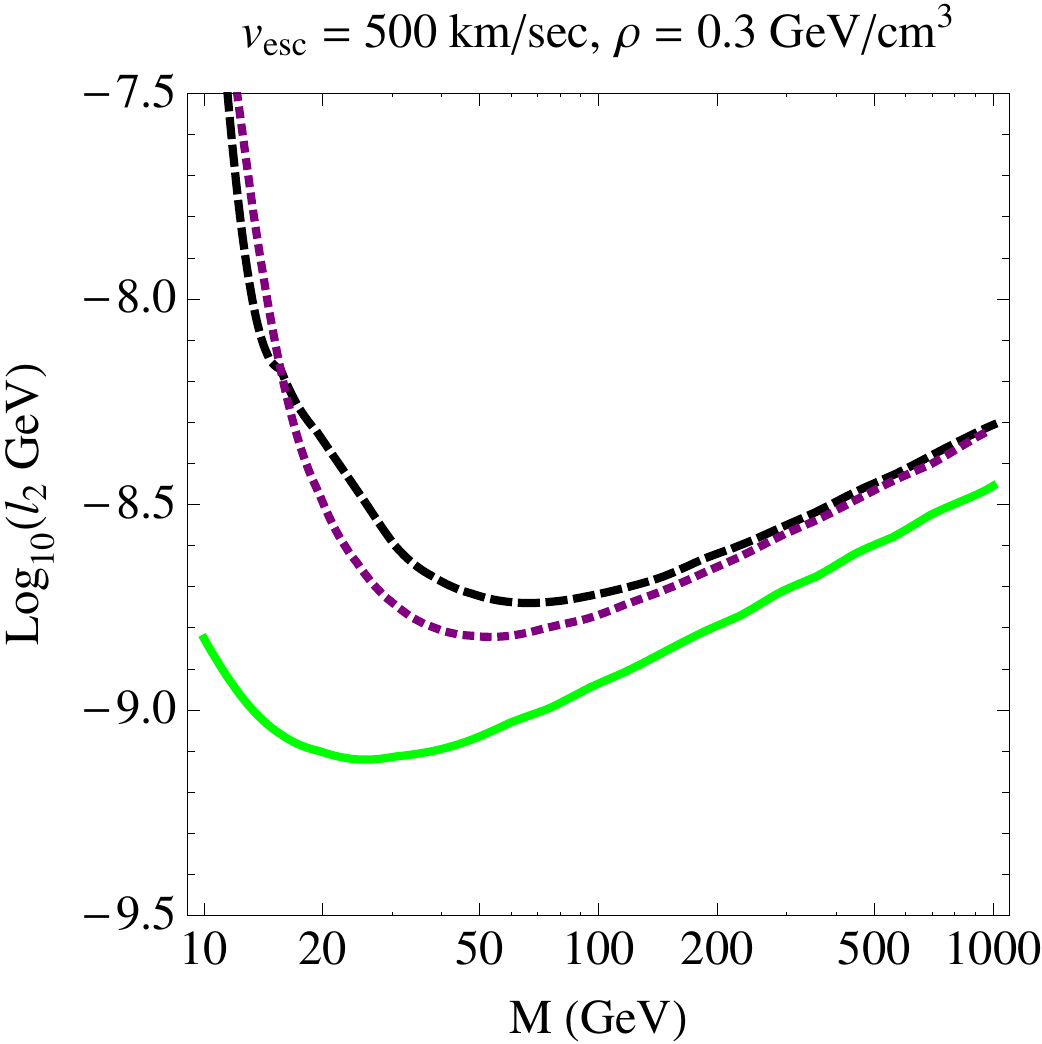}& \\
\includegraphics[scale=0.6]{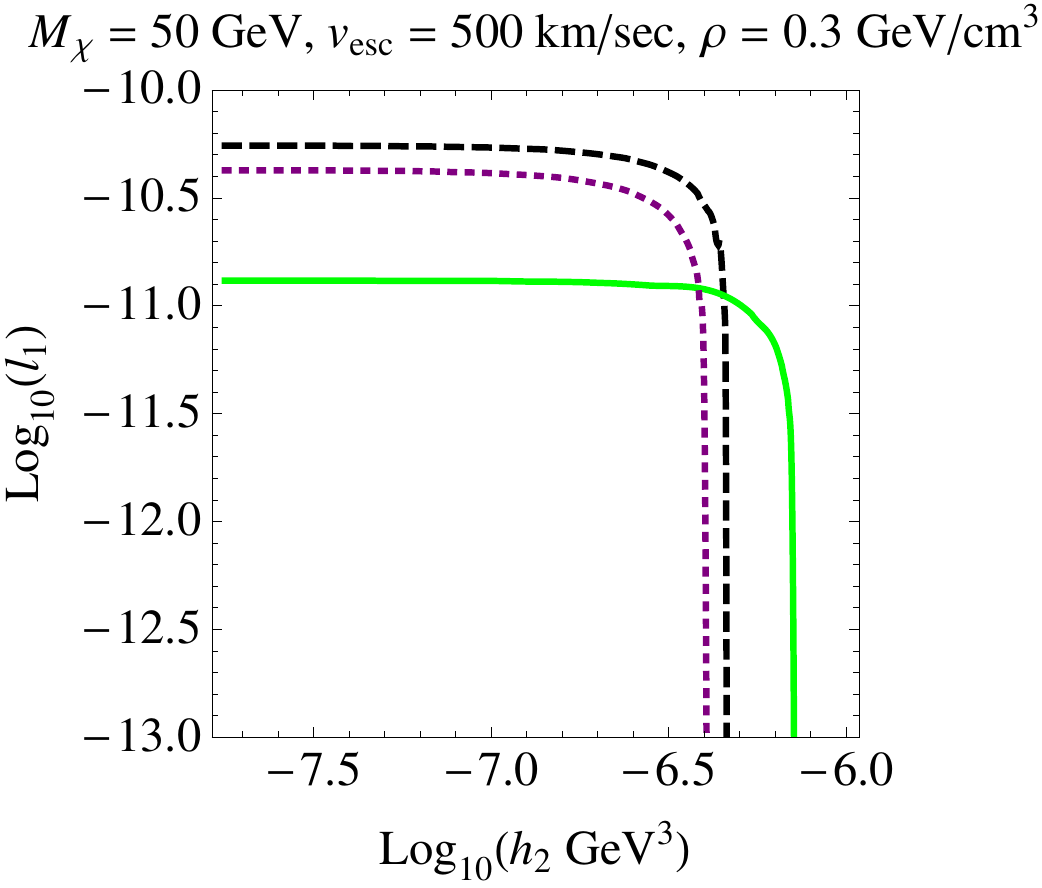}& 
\end{tabular}
\caption{SI exclusion curves for Wilson coefficients with CDMS black dashed, Xenon10 green solid, Xenon100 purple dotted. }
\label{fig: SI constraint}
}

\FIGURE[p!]{
\begin{tabular}{cc}
 \includegraphics[scale=0.5]{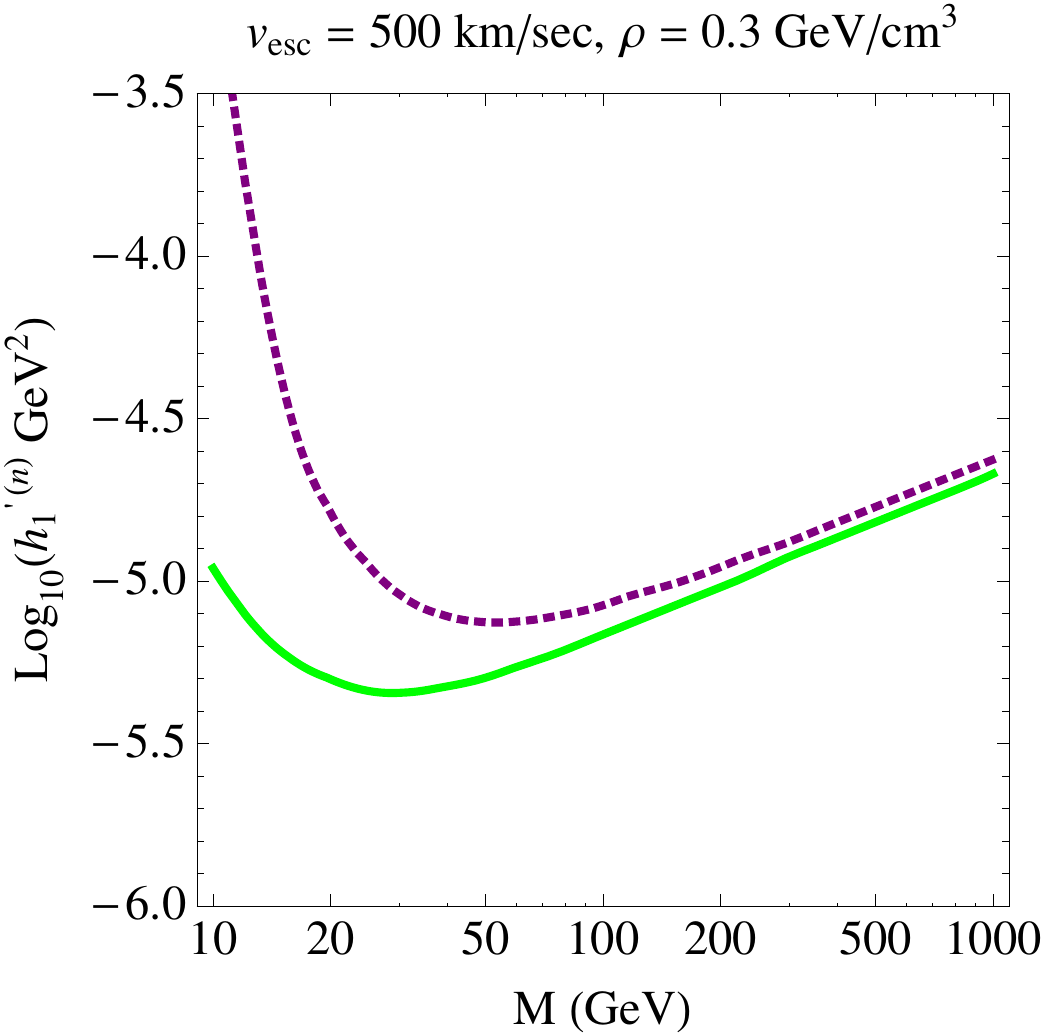}&
\includegraphics[scale=0.5]{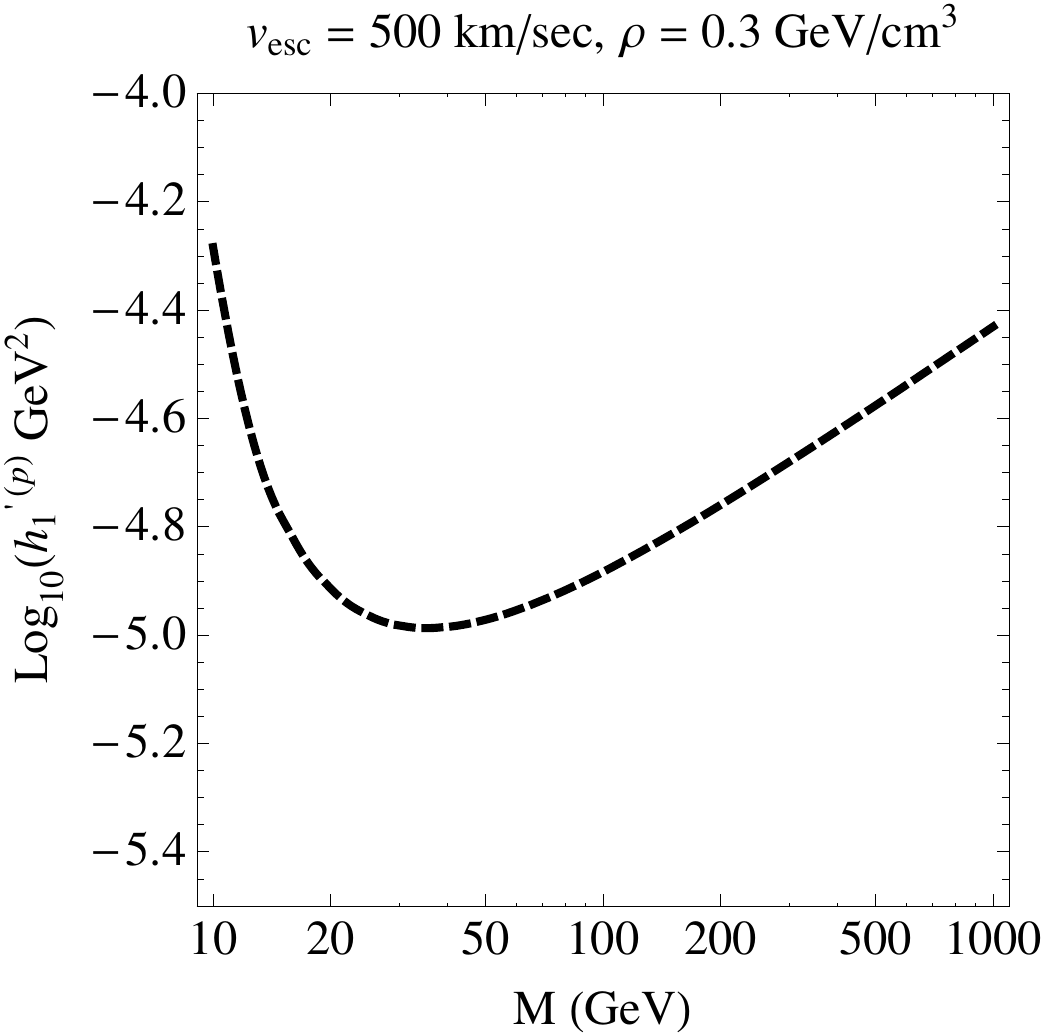}\\
 \includegraphics[scale=0.5]{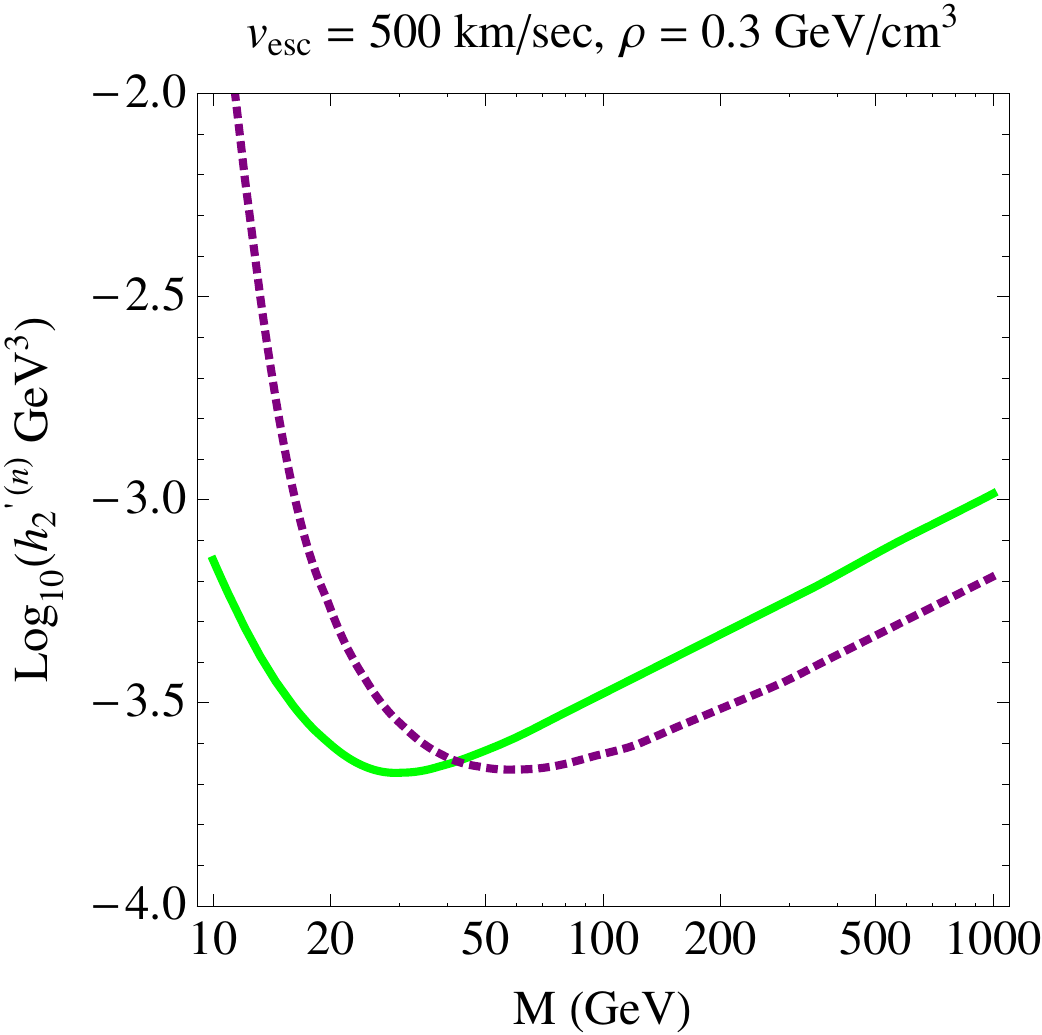}&
\includegraphics[scale=0.5]{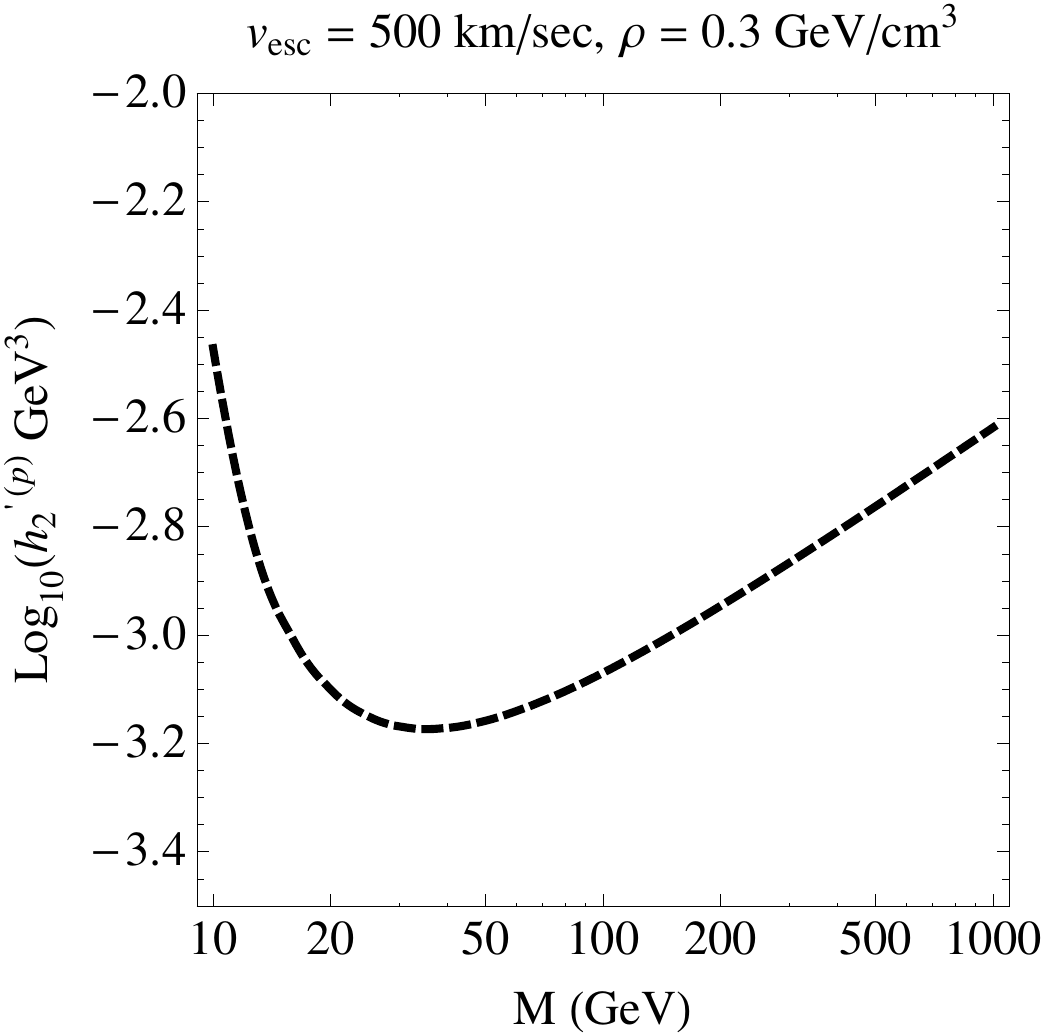}\\
\includegraphics[scale=0.5]{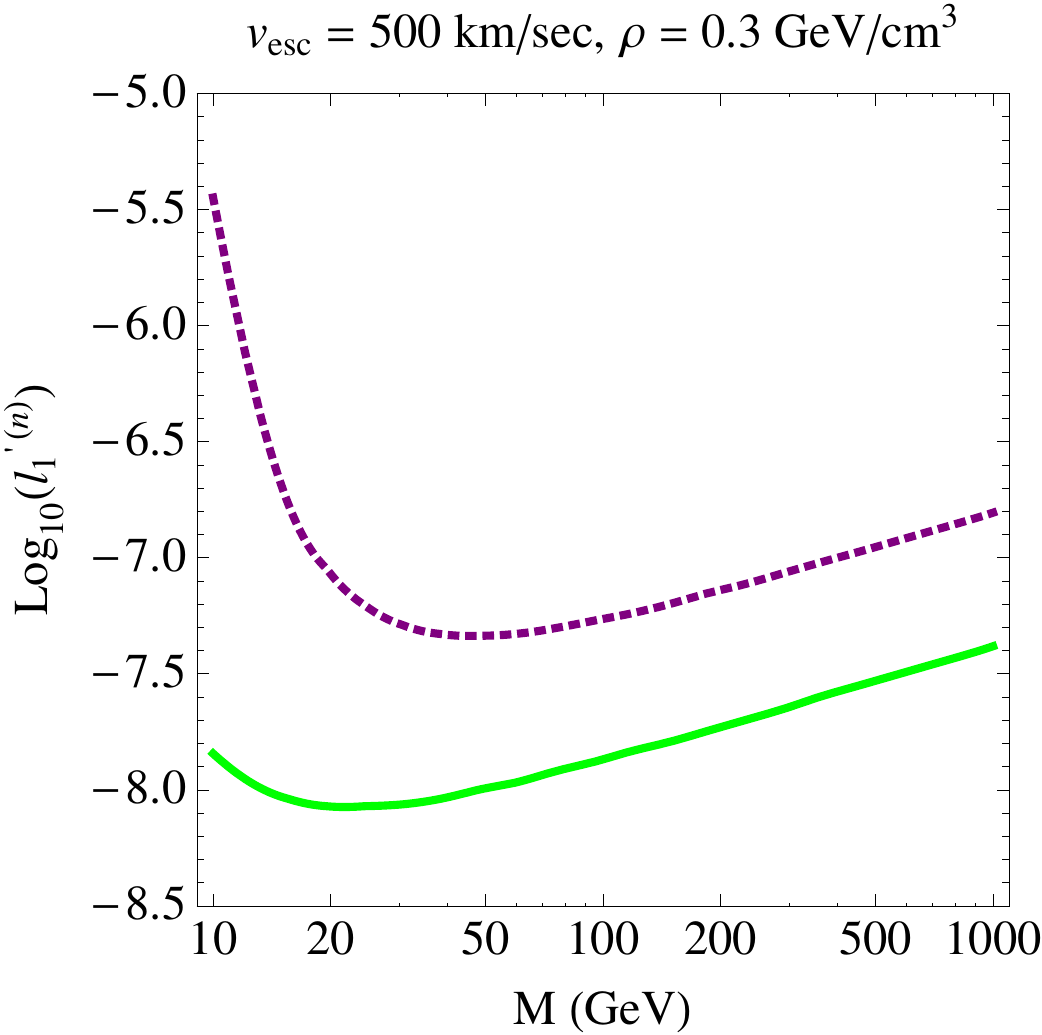}&\includegraphics[scale=0.5]{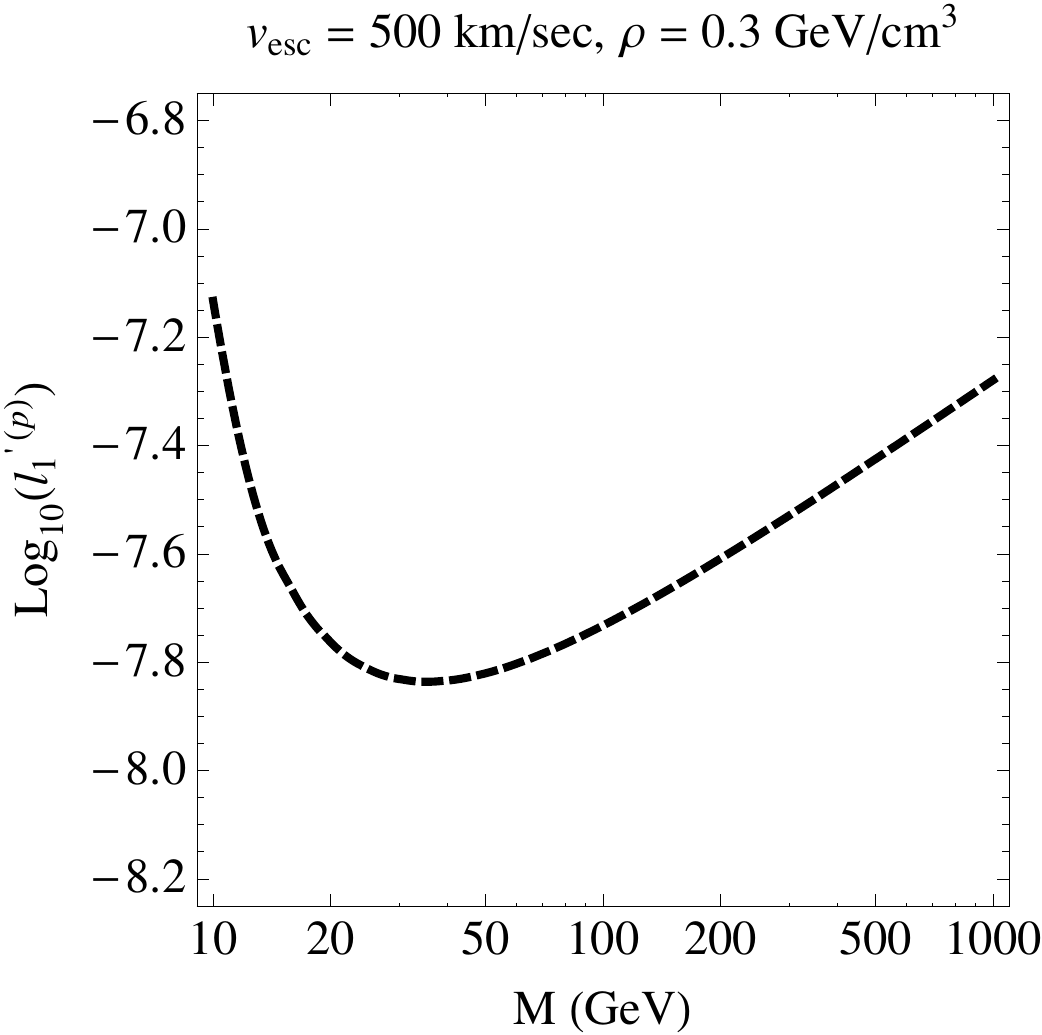}\\
\includegraphics[scale=0.5]{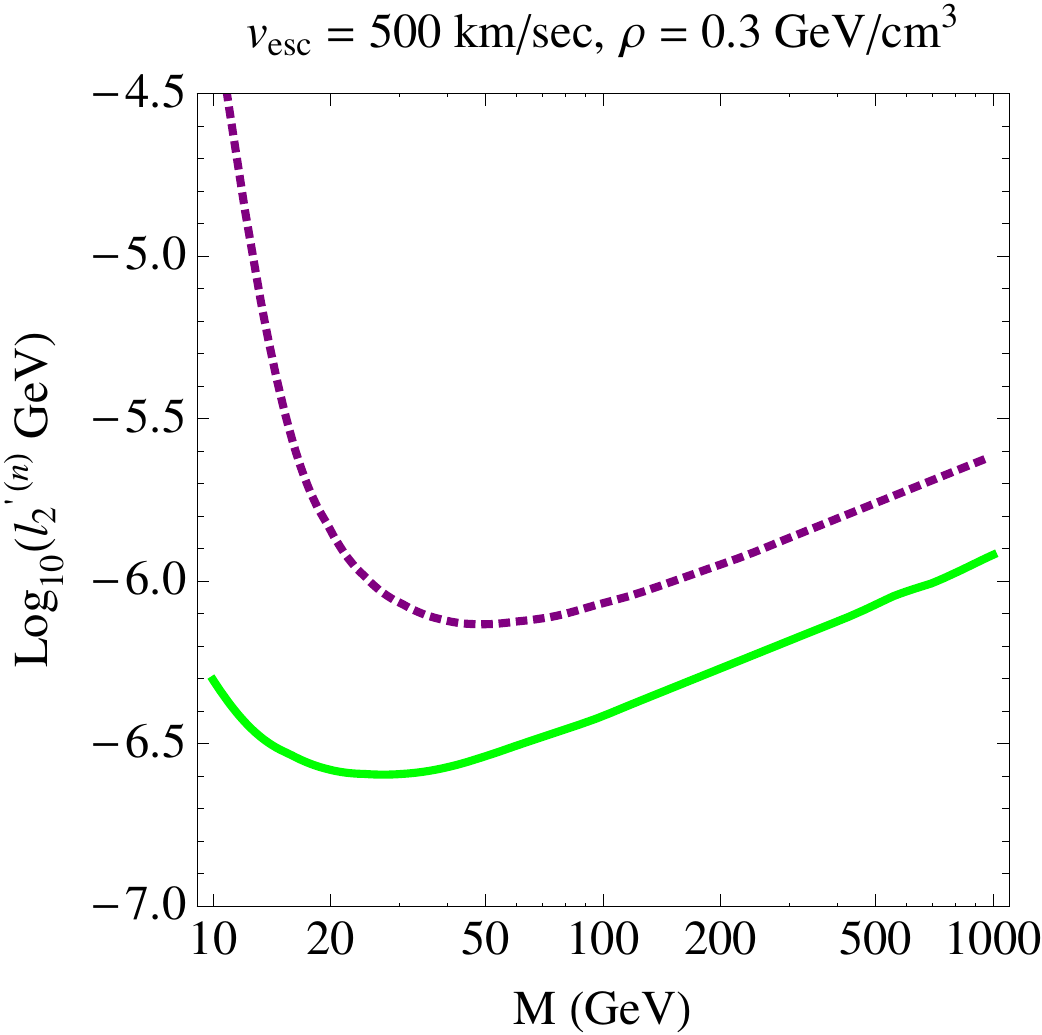}&\includegraphics[scale=0.5]{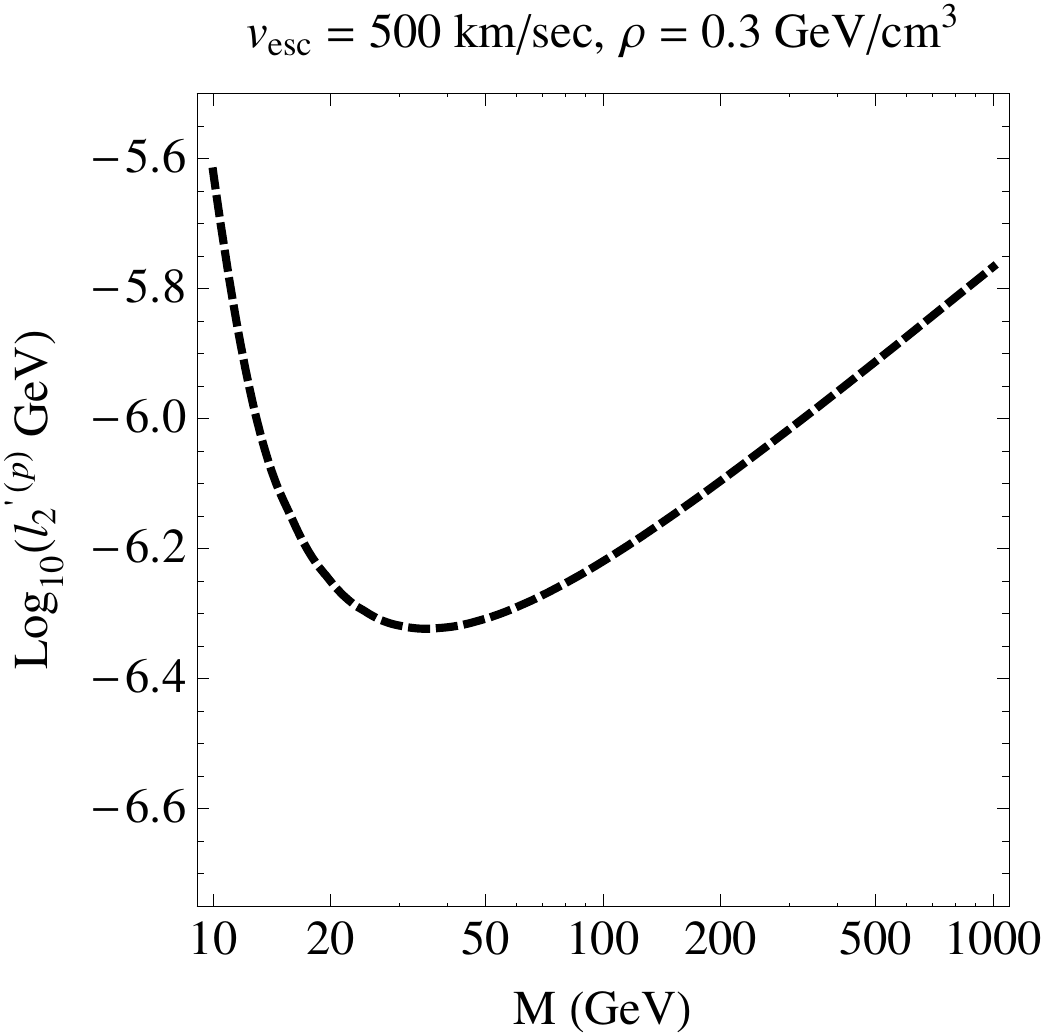}
\caption{SD exclusion curves for Wilson coefficients with Xenon10 green solid, Xenon100 purple dotted (DM--neutron), COUPP black dashed (DM--proton).}
\end{tabular}
\label{fig: SD constraint}}

The limits on each coefficient from the experiments we consider are displayed in Fig.~\ref{fig: SI constraint} and Fig.~\ref{fig: SD constraint}. Notice that in cases with a light mediator, due to the enhanced rate at small recoil energy, the XENON10 reanalysis from Ref. \cite{Angle:2009xb} becomes the strongest constraint, as it is the result with the lowest energy threshold. So far the importance of energy calibration around the threshold has been mostly emphasized for ruling in or out the light DM scenario \cite{Collar:2010gg}. For Xenon-based experiments, the recoil energy threshold is determined by the scintillation efficiency ${\cal{L}}_{eff}$, which is still controversial at the moment~\cite{Aprile:2010um, Collar:2010gg}. But as is clear from the discussions above, the low energy region and the precise measurement of ${\cal{L}}_{eff}$ are also crucial for constraining heavy DM with a light mediator. 
From Fig.~\ref{fig: SI constraint}, the strongest constraints on the coefficients of SI operators in Eq.~(\ref{eq: Hamiltonian}) are 
\beqs
h_1 & \simlt & 10^{-8}\,{\rm{GeV}}^{-2} = \frac{10^{-4}}{\left(100\, \rm{GeV}\right)^2}\nonumber \\
h_2 & \simlt & 10^{-7}\,{\rm{GeV}}^{-3} = \frac{10^{-1}}{\left(100\, \rm{GeV}\right)^3}\nonumber \\
l_1 & \simlt & 10^{-11}\nonumber\\
l_2 & \simlt & 10^{-9}\,{\rm{GeV}}^{-1} = \frac{10^{-7}}{\left(100\, \rm{GeV}\right)}.
\label{eq: si constraints}
\eeqs
We have written the couplings as a ratio $\frac{g}{\Lambda^k}$, where the dimensionless part $g \sim g_\chi f_{p(n)}$ is a product of the strength of the DM coupling to the mediator and the nucleus coupling to the mediator, and we have chosen the scale involved, $\Lambda$, to be 100 GeV for illustrative purposes. (Parametrically, $\Lambda$ could be $m_\chi, m_N, m_\phi$, or some new scale in the dark sector). From this decomposition, one can draw clues about the relevant scales and couplings of a model that could give rise to the NR operators. For instance, as already discussed in Sec.~\ref{sec: models}, the simplest contact interaction could come from Higgs exchange between DM and nucleus with Higgs mass around 100 GeV and its coupling to nucleus $\calo (10^{-3} -10^{-4})$. We also include in Fig.~\ref{fig: SI constraint} an exclusion plot in ($h_2,l_1$) plane for 50 GeV DM.

We plot constraints on the SD operators in Fig. \ref{fig: SD constraint}. We have assumed purely neutron or purely proton couplings, for simplicity. In the case of the DM--neutron coupling, bounded by XENON10 and XENON100, the analysis is complicated by the fact that there are two important isotopes, $^{129}$Xe and $^{131}$Xe, with different spin interactions. We assume that these couplings scale as universal numbers $h_1^{\prime(n)}, h_2^{\prime(n)}, l_1^{\prime(n)}, l_2^{\prime(n)}$ times the isotope-dependent factor $\left<S_n\right>/J$:
\beq
h_1^\prime(^{A}{\rm Xe}) = h_1^{\prime(n)}\left.\left(\frac{\left<S_n\right>}{J}\right)\right|_{^A{\rm Xe}},~~~~(A = 129,131)
\eeq
and similarly for the other couplings. For $^{129}$Xe, $J = 1/2$ and we take $\left<S_n\right> = 0.359$; for $^{131}$Xe, $J = 3/2$ and we take $\left<S_n\right> = -0.236$ \cite{Angle:2008we}. We plot limits on $h_1^{\prime(n)}$, etc., rather than on the isotope-dependent couplings $h_1^\prime$. The DM--neutron coupling strongest bounds are
\beqs
h_1^{\prime(n)} & \simlt & 10^{-6}\,{\rm{GeV}}^{-2} = \frac{10^{-2}}{\left(100\, \rm{GeV}\right)^2}\nonumber \\
h_2^{\prime(n)} & \simlt & 10^{-4}\,{\rm{GeV}}^{-3} = \frac{10^{-1}}{\left(10\, \rm{GeV}\right)^3}\nonumber \\
l_1^{\prime(n)} & \simlt & 10^{-8}\nonumber\\
l_2^{\prime(n)} & \simlt & 10^{-7}\,{\rm{GeV}}^{-1} = \frac{10^{-5}}{\left(100\, \rm{GeV}\right)}.
\label{eq: sd constraints}
\eeqs
In the DM--proton case, no similar ambiguity arises, as we assume the scattering at COUPP is entirely on protons in the single isotope $^{19}{\rm F}$. Nonetheless, in order to have roughly comparable bounds in the two cases, we define $h_1^{\prime(p)}$ by $h_1^\prime = h_1^{\prime(p)} \left.\left(\left<S_p\right>/J\right)\right|_{^{19}{\rm F}}$, and similarly for the other couplings.
The DM--proton coupling constraints are a little bit weaker than the DM--neutron constraints. Notice that for the momentum-suppressed SD contact interaction to be detectable, it is necessary to have either a large coupling or the scales involved have to be small $\lesssim \calo$ (10$\, $GeV).

\subsection{Constraints on force carriers from other experiments }
\label{sec: other constraints}
In this section we briefly comment on constraints from other experiments on the mediators. Many constraints are model-dependent and mild compared to the direct detection constraints. A complete discussion is beyond the scope of this paper. 

First we consider constraints on heavy mediators from colliders. The interaction between the DM and quarks and gluons will also lead to direct DM production at hadron colliders such as the Tevatron. With initial state radiation, this may lead to mono-jet signals. Recently there were two studies of constraints on the contact interaction between DM and colored particles from monojet $+$ missing $E_T$ searches at the Tevatron~\cite{Goodman:2010yf, Bai:2010hh}. They have shown that for a heavy mediator with $m_\phi >$ 100 GeV and SI interaction, collider searches have greater reach $\sigma \sim 10^{-40}$ cm$^2$ for light DM with $m_\chi <$ 10 GeV, to which direct detection experiments have limited sensitivity. But for $m_\chi >$ 10 GeV, direct detection has a more stringent bound. For SD interactions, the situation is more interesting. The Tevatron bound on cross section of contact interaction is $(10^{-39} - 10^{-40})$ cm$^2$ for $m_\chi <$ 200 GeV, one or two orders of magnitude below that from direct detection. But once the mediator mass drops below 100 GeV, the production process is no longer from a contact interaction in the collider. The collider constraints weaken and the direct detection experiments have stronger or comparable bounds for $m_\chi >$ 10 GeV for both SI and SD interactions.

For a light mediator with mass around or below a GeV, there could be model dependent constraints on its coupling to quarks from rare meson decays. Depending on the flavor structure of the couplings, rare Upsilon and kaon decays could be relevant. For example, $\Upsilon \to \gamma \phi$ constrains the couplings to the third generation of quarks and hence couplings to gluons (see Sec.~\ref{sec: relativistic matching} for the relation between couplings to heavy quarks and gluons).  If the mediator couplings to the first two generations of quarks are not flavor diagonal, e.g. if $l_1\bar{d}s\phi$ exists, there will be constraints from $K^+  \to \pi^+ \phi$. Precise bounds depend on the final state of $\phi$ decays. For instance, if $\phi \to$ invisible, the experimental bound on the kaon branching ratio $Br(K^+  \to \pi^+ +$ invisible) $< 7.3 \times 10^{-11}$ gives $l_1< 10^{-12}$~\cite{Deshpande:2005mb}.\footnote{Pure pseudoscalar coupling does not mediator this process unless $\pi, \phi$ mixes.} More constraints from meson decays for vector mediators can be found in~\cite{Fayet:2006sp}. 

While mediator couplings to leptons are free parameters in general, if they are nonzero, they will be constrained by $g-2$ measurement, beam-dump experiments and astrophysical constraints. First, the $g-2$ experiment rules out large couplings to the leptons above $10^{-3}$. For gauge boson mass below $m_\mu$, $\delta a_\mu$ from axial vector coupling has an additional enhancement factor from the propagator, $m_\mu^2/m_\phi^2$, compared to $\delta a_\mu$ from other bosons and couplings~\cite{Fayet:2007ua}. In this case, the bound gets stronger with the coupling $c \lesssim 10^{-6} m_{\phi}/$MeV. The beam-dump experiments exclude couplings of order $10^{-3}-10^{-7}$ depending on the mediator mass. There are still small allowed islands in the parameter space around $m_\phi \sim$ (10 - 100) MeV and $c \sim$ $(10^{-5} - 10^{-7})$. Detailed analysis could be found in~\cite{Bjorken:2009mm, Bjorken:1988as}. For even smaller couplings, the constraints are from supernova cooling.  If the light particle decays $\gtrsim$ 10 km from the point of production inside the supernova, it will contribute to the energy loss. To be compatible with the observed SN 1987 pulse duration, the new channel's contribution to the energy loss is estimated to be smaller than $10^{53}$ erg/s. For vector mediator with mass below 100 MeV, this requirement rules out couplings around $10^{-6} - 10^{-8}$~\cite{Bjorken:2009mm}. For axion-like mediator with $m_\phi \sim$ (1 - 200) MeV, the excluded coupling is of order $10^{-9}$~\cite{Jesse}.

\afterpage{\clearpage}

\section{Nuclear physics and form factors}
\label{sec:nuclear}

There is a subtlety involved in describing DM direct detection in terms of an effective field theory. The momentum transfers involved are $q \sim 100$ MeV, so $1/q$ can be of the same order as nuclear radii $R \sim A^{1/3}$ fm (recall that 1 fm$^{-1} \approx 200$ MeV). This implies that the nuclei are not pointlike from the perspective of DM, and we should not, strictly speaking, describe DM--nucleus interactions in terms of contact interactions $V(r) \sim \delta^3(\vec{r})$; the DM--nucleus cross section will involve a form factor $|F(q^2)|^2$.

Other aspects of nuclear physics can also influence the treatment of NR effective field theories. We have three small parameters, $v_{DM}, q/M_\chi,$ and $q/M_N$, all of which are of order 10$^{-3}$. However, nuclear binding energies are of order 10 MeV per nucleon, so inside the nucleus, individual nucleons have velocities $v_n \sim 10^{-1}$. One might worry that this, along with $q/M_n$, is a large expansion parameter that could spoil the NR effective theory. We will review the treatment of nuclear form factors in DM scattering to argue that this does not occur.

We would also, in the spirit of effective field theory, like to give an overview of the various operators that can be involved in nuclear scattering. A full review of those aspects of the physics of nucleons and nuclei that are relevant for DM direct detection is beyond the scope of this paper. Nonetheless, we will point out some lacunae in the DM literature. In some cases, we will offer pointers into the QCD and nuclear literature that can begin to fill these gaps. The early direct detection literature understandably focused on the operators $\bar{q}\gamma_\mu \gamma_5 q$ and $\bar{q}q$ that are most relevant for direct detection in the MSSM, linking their matrix elements to the nucleon spin and the pion-nucleon sigma term \cite{Goodman:1984dc,Ellis:1991ef,Engel:1991wq}. The nuclear physics involved in these operators has been reviewed many times \cite{Engel:1992bf,Jungman:1995df,Lewin:1995rx}. More recent literature has broadened the set of operators considered, but often a discussion of nuclear matrix elements has been omitted. 

\subsection{Matching to relativistic operators}
\label{sec: relativistic matching}

First, let us discuss the most basic step of the matching, from a renormalizable QFT to effective operators of the form $J_{q(g)} J_\chi$. Suppose we integrate out a heavy mediator, say a scalar of mass $M$ exchanged between DM and a quark. The leading term will be a contact interaction $\bar{\chi}\chi\bar{q}q$ for which we care about the form factor of a scalar current $\bar{q}q$ in the nucleus. What about subleading terms? The full tree-level amplitude is $\bar{\chi}\chi \frac{1}{q^2 - M^2} \bar{q} q$, which has an expansion in $\frac{q^2}{M^2}$ but in which every term will depend on the same form factor $\left<N| \bar{q}q|N\right>$. When integrating out the heavy scalar, one could also draw diagrams with radiated gluons, but this generates operators suppressed by both factors of $\alpha_s(M)$ and further powers of $M$, and we can safely neglect them. (In certain cases one may wish to include corrections involving derivatives, as in the twist-two coupling $\bar{\chi}\gamma^\mu\partial^\nu \chi {\cal O}^{(2)}_{q\mu\nu}$ studied in \cite{Drees:1993bu}. In any case, one expects that in general operators with many derivatives are unimportant.)

Next one can consider again the case of exchange of a particle between DM and a quark, but in the case that this particle is light, so that the amplitude is $\bar{\chi}\chi \frac{1}{q^2} \bar{q}q$. Again the form factor is $\left<N| \bar{q}q|N\right>$. Unlike our previous argument, one cannot now argue that radiating additional gluons is suppressed by small $\alpha_s$ or powers of the mediator mass $M$. However, one doesn't have to: in this case, we are not integrating out a particle and matching onto local operators at a high scale $M$, so we should never think of the quarks as perturbative. We simply have a process with amplitude $\bar{\chi}\chi \frac{1}{q^2} \bar{q}q$ in the low-energy theory, and we are evaluating the matrix element $\left<N| \bar{q}q|N\right>$ in the nonperturbative QCD regime. Any gluons radiated from the quarks should be thought of as part of the nonperturbative ``blob" connecting the local operator $\bar{q}q$ to the nucleon external states.

Another possibility is that we have an $s$-channel particle, like a squark in a process $\chi q \to \tilde{q}^* \to  \chi q$. In this case the amplitude has the form $\chi\bar{q} \frac{1}{s - M^2} \bar{\chi} q$, so the $\bar{q}$ and $q$ are not at the same point and in principle we care about a form factor for a {\em nonlocal} operator, which might be thought of as involving contributions of infinitely many local operators. However, we are saved by the fact that $s \sim (m_N + m_\chi)^2 \gg (q^2, \Lambda_{QCD}^2)$, so that for all practical purposes the operator we generate may again be treated as local (up to $q^2/s$-suppressed terms). We can Fierz $(\chi \bar{q})(\bar{\chi}q)$ to obtain an expression in terms of local gauge-invariant operators like $\bar{q}q$ and $\bar{q}\gamma^\mu q$, and then use the simple form factors $\left<N| \bar{q}\Gamma q|N\right>$. Note that this conclusion is independent of the mediator mass $M$, provided the process is not resonant with $s \approx M^2$. In particular, even though one cannot strictly ``integrate out" a {\em light} $s$-channel particle, for the purpose of understanding nuclear form factors we can still treat this case as leading to a local operator $\bar{q} q$.

DM can interact not only with quarks but with glue. The simplest way to generate an operator like $\bar{\chi}\chi G^a_{\mu\nu}G^{a\mu\nu}$ is if DM exchanges a scalar (like the Higgs) with heavy quarks. Alternatively, DM could exchange a pseudoscalar with heavy quarks. In both cases we can integrate out the heavy quarks to produce effective couplings to gluons. The matching is \cite{SVZ}:
\beqs
m_Q \bar{Q}Q & \to & -\frac{2}{3} \frac{\alpha_s}{8\pi} G^{a}_{\mu\nu}G^{a\mu\nu} \\
m_Q \bar{Q} \gamma_5 Q & \to & i \frac{\alpha_s}{16\pi} G^a_{\mu\nu}\tilde{G}^{a\mu\nu}.
\eeqs
It is natural to then ask what happens when integrating out heavy quarks exchanging other types of mediators. This possibility has received relatively little attention in the DM context, but most of the ingredients necessary to understand it exist in the QCD literature. The background field method gives an efficient technique for computing the effective gluonic operators generated in such cases \cite{Novikov:1983gd}. For instance, a new vector coupling to a heavy-quark current like $\bar{Q} \gamma^\mu Q$ must match to some conserved current below the scale $m_Q$. One generally expects that conserved currents are dimension 3, but there is another possibility: they can be descendants of other operators, rather than primaries, so that $J_\mu$ matches on to an operator of the form $\partial_\nu \tau_{\mu\nu}$ with $\tau_{\mu\nu}$ an antisymmetric tensor. This must be the case in order to match to a gluonic operator after integrating out the heavy quark. The result is \cite{KaplanManoharStrange}:
\beqs
\bar{Q} \gamma_\mu Q & \to & \frac{g_s^3}{(16\pi^2)^2~45 m_Q^4} \left(5 \partial_\alpha {\rm Tr}G_{\sigma \tau}\{G^{\sigma \tau},G_{\alpha \mu}\}-14 \partial_\alpha {\rm Tr} G_{\mu\sigma}\{G^{\sigma \tau}, G_{\tau \alpha}\}\right) \label{eq:heavyvector} \\
\bar{Q} \gamma_\mu \gamma_5 Q & \to & \frac{g_s^2}{48\pi^2 m_Q^2} \epsilon_{\mu\rho\tau\sigma} {\rm Tr}\left(G^{\alpha\rho}\partial_\alpha G^{\tau \sigma} + 2G^{\tau\sigma}\partial_\alpha G^{\alpha \rho}\right). \label{eq:heavyaxial}
\eeqs
On the other hand, if we have a dipole-type interaction with $\bar{Q}\gamma_5 \sigma_{\mu \nu} Q$, then integrating out the heavy quark generates couplings to dimension-six gluonic antisymmetric tensor operators. Again the matching is best calculated using the background field method \cite{DeRujula:1990wy,Chang:1992vs}:
\beq
\bar{Q}\gamma_5 \sigma_{\mu \nu} Q \to \frac{g_s^3}{192\pi^2 m_Q^3} d_{abc} \tilde{G}^a_{\rho\sigma}G^{b\rho\sigma}G^{c}_{\mu\nu}.
\eeq
These operators are suppressed by powers of the heavy-quark scale, so we expect that they would be overwhelmed by contributions from light-quark currents except in models where DM interacts primarily with the third generation through new forces. It would be interesting, but beyond the scope of this paper, to consider how the size of effects in direct detection and flavor physics might be related in such models.

These arguments convince us that, despite the in-principle issue that DM--nucleus scattering can depend on arbitrary form factors $\left<N(p+q)|{\cal O}(q)|N(p)\right>$, in practice only a handful of form factors (for operators made up of two quarks or two or three gluonic field strengths) will be relevant and contributions of others will be suppressed by powers like $q^2/M^2$ or $q^2/m_N^2$, sometimes with small perturbative couplings $\alpha_s$ in front.

\subsection{Matching quark and gluon operators to nucleon operators}
\label{sec: match to nucleon}

Once we have an effective theory involving DM currents coupled to quark or gluon currents, the next step is to match onto a theory of nucleons. The justification for this is that we consider momentum transfers too small to resolve the structure of an individual nucleon. This matching is accomplished by considering the matrix element of the QCD operator between nucleon states.

For the scalar operators $\bar{q}q$, the result is well-known to be related to the so-called pion-nucleon sigma term \cite{TPCheng,HYCheng}. The nucleon mass arises from the trace anomaly:
\beq
M_n = \left<n | \theta^\mu_\mu |n\right> = \sum_q m_q \left<n | \bar{q}q | n\right> + \frac{\beta(\alpha_s)}{4\alpha_s} \left<n | G^a_{\mu\nu}G^{a\mu\nu}|n\right>,
\eeq
so one defines the fraction of nucleon mass arising from the various quark flavors and the gluons as:
\beqs
f^{(n)}_{T_q} & = & \frac{\left<n|m_q \bar{q}q|n\right>}{m_n}, \\
f^{(n)}_{TG} & = & 1 - \sum_{q=u,d,s} f^{(n)}_{T_q}.
\eeqs
Thus both the scalar quark operators $\bar{q}q$ and the scalar gluon operator ${\rm Tr}G^2$ match onto the scalar nucleon operator $\bar{n}n$ with coefficients that depend on how the various quark flavors contribute to the nucleon mass. A recent discussion of uncertainties in these quantities appears in \cite{EllisOliveSavage}.

The vector current case is easy to understand because we have a conserved current:
\beq
\bar{q} \gamma_\mu q \to Q_n \bar{n} \gamma_\mu n,
\eeq
where $Q_n$ is the charge of the nucleon under the corresponding current. In particular, only valence quarks contribute at leading order in this case, not sea quarks or gluons. On the other hand, heavy-quark currents can lead to {\em dipole} moments even though the nucleon has no net charge under these currents; such dipole moments have been discussed in  \cite{Ji:2006vx}. A subtlety arises in computing them. From eq. \ref{eq:heavyvector} one might naively expect an $m_Q^{-4}$ suppression of the matrix element. However, the 3-gluon operator we matched to was renormalized at a scale $m_Q$, and its matrix element in the nucleon has a quadratic divergence, so that the dipole moment is in fact suppressed only by $m_Q^{-2}$.

The next familiar case is the axial-vector current $\bar{q}\gamma_\mu\gamma_5 q$, which is related to the nucleon spin \cite{Goodman:1984dc}. The matching is:
\beq
\bar{q} \gamma_\mu \gamma_5 q \to \Delta_q \bar{n}\gamma_\mu \gamma_5 n,
\label{eq:spinmatching}
\eeq
where $\Delta_q$ is the fraction of the spin of the nucleon carried by quark flavor $q$. These fractions are given by integrals of helicity-dependent parton distributions, as reviewed in \cite{JaffeReview}, which are well-measured \cite{EuropeanMuonCollaboration, Airapetian:2007mh,Alekseev:2007vi}. The case of the pseudoscalar current $\bar{q} \gamma_5 q$ is related by PCAC (or a Goldberger-Treiman relation) to that of the axial current \cite{HYCheng}. For the axial current of a heavy quark, we match to the operator involving gluons, as in eq. \ref{eq:heavyaxial}. The matrix element of these gluonic operators has been argued to be related to a higher-twist matrix element that is known (and small) \cite{Polyakov:1998rb}.

The tensor operator $\bar{q} \sigma_{\mu \nu} q$ is of interest if the nucleus has a dipole interaction with some new gauge boson. This operator has appeared in the DM literature, often in the context of a general operator analysis (including, but not necessarily limited to, refs. \cite{Kurylov:2003ra, Kopp:2009qt,Agrawal:2010fh}), but to the best of our knowledge the only correct discussion in the DM context of its matrix element in the nucleon is in ref. \cite{Belanger:2008sj}. This operator is familiar in the spin physics context, and is related to the ``transversity distribution" $h_1(x)$ of the nucleon \cite{RalstonSoper, JaffeJi}. In particular, although the NR limit of the operators $\bar{\psi}\gamma_\mu\gamma_5 \psi$ and $\bar{\psi}\sigma_{\mu\nu}\psi$ agree, in the relativistic limit they measure {\em different} quantities, and the matching of quark-level operators to nucleon-level operators is different in the two cases. The analogue of the spin fraction in eq. \ref{eq:spinmatching} is the ``tensor charge" $\delta_q$:
\beq
\bar{q} \sigma_{\mu\nu} \gamma_5 q \to \delta_q \bar{n}\sigma_{\mu\nu} \gamma_5 n.
\label{eq:tensormatching}
\eeq
Despite being twist-two parton distribution functions on an even theoretical footing with the well-studied cases, the transversity distributions are more difficult to measure and relatively poorly known. The tensor charges for light quarks in the proton have only recently been extracted from experimental data \cite{TransversityExperiment}, giving, at $Q^2 = 0.8$ GeV$^2$,
\beqs
\delta_u & = & 0.54^{+0.09}_{-0.22} \\
\delta_d & = & -0.23^{+0.09}_{-0.16}. 
\eeqs

\subsection{Nuclear form factors for nucleonic currents}
\label{sec: nucleon currents}

Now that we have matched our effective theory onto operators involving nucleons, it remains to evaluate the correlators of nucleonic currents in nuclei. For instance, we may wish to calculate $\left<N(p+q)|\bar{n}n(q)|N(p)\right>$. It is at this stage that we can take an NR limit and view this as simply a distribution of the number of nucleons or the spin of nucleons within a nucleus. These are familiar form factors \cite{Engel:1992bf, Lewin:1995rx}; see refs. \cite{Duda:2006uk, Bednyakov:2006ux} for more recent discussions of the status of our knowledge of the SI and SD form factors in various isotopes.

In plotting limits we have used the usual SI form factor $F^2(E_R)$ and SD form factor $S(E_R)/S(0)$. In general, however, the story is more elaborate. Note that at this stage of the matching, multiple form factors can arise for a given process, e.g. the familiar decomposition for a coupling to a conserved current:
\beq
\left<N(p+q)| \bar{n}\gamma^\mu n(q) | N(p)\right> = \bar{N}(p+q) \left(F_1(q^2) \frac{(2p+q)^\mu}{2m_N} + \left(F_1(q^2) + F_2(q^2)\right) \frac{i \Sigma^{\mu\nu} q_\nu}{m_N}\right) N(p).
\eeq
Here we have suppressed Lorentz indices on the spin-$J$ nucleus wavefunctions $N(p)$, which have the relativistic normalization (e.g. $u_N(p)$ for spin-1/2 nuclei), and $\Sigma_{\mu\nu}$ is the spin-$J$ Lorentz representation (e.g. $\frac{1}{2} \sigma_{\mu\nu}$ for fermions). It is only at this stage, when we have an effective coupling to the entire nucleus, that we can take NR limits of the spinor structure, finding that the $\gamma^i$ term is suppressed by the nucleus velocity relative to the $\gamma^0$ term measuring the total charge. In particular, note that the nucleon velocity $v_n \sim 0.1$ does {\em not} appear, as it is not possible to take an NR limit at an earlier stage of the matching process.

The second, magnetic dipole moment, term is suppressed by $q/m_N \approx 10^{-3}$ relative to the leading term, assuming $F_2(q^2)$ is not much larger than $F_1(q^2)$.  However, depending on which DM operator $\bar{q}\gamma^\mu q$ is contracted with, it is possible that the two terms contribute in comparable ways. This possibility has recently played a role in several very recent attempts to understand whether data from DAMA, CoGeNT, and other experiments are compatible \cite{Chang:2010en, Barger:2010gv, Fitzpatrick:2010br, Banks:2010eh}. If the coupling is to $\bar{\chi}\sigma_{\mu\nu} \chi q^\nu$, then in the NR limit the dominant part of the DM operator is $\sim (\vec{s}_\chi \times \vec{q})$, which is dotted into the velocity suppressed part of the quark current $\sim Z\vec{v} F_1(q^2)$. On the other hand, the $\gamma^0$ term in the quark current contract with a velocity suppressed part of the DM operator. Thus $F_1(q^2)$ multiplies the operator $\vec{s}_\chi \cdot (\vec{v} \times \vec{q})$ in the NR limit. However, the magnetic moment term can be contracted without extra velocity suppression, corresponding to the operator $F_2(q^2) q^2 \vec{s}_\chi \cdot \vec{s}_N$ in the NR limit. In this way, the $F_1(q^2)$ and $F_2(q^2)$ contributions can be comparably important in certain models. 

\section{Conclusion and outlook}
\label{sec: conclusion}
In this paper we have constructed a simple NR effective theory to study signals from DM direct detection experiments. Different operators in the NR theory correspond to different types of interactions between DM and SM quarks. They lead to qualitatively different recoil spectra. Thus if DM is discovered in the near future DM direct detection, the recoil spectrum will constrain the NR effective theory and its possible field theory completions. Valuable information on the nature of DM--nucleus interactions can be obtained. 

There are still many details that need to be studied to optimize our understanding of possible DM signals. For instance, both the nuclear form factor and the DM velocity distribution could modify the recoil spectrum and smear the differences from DM dynamics. As far as we know, a complete list of the matching between quark or gluon operators and nucleon operators and the nuclear form factors is still absent in the literature. Another issue comes from the DM velocity distribution. In this paper we used the traditional simplest Maxwellian distribution. But the DM velocity distribution in our galaxy is certainly more complicated, and non-Maxwellian distributions may result in more structures in the recoil spectrum. It will be interesting to develop methods to disentangle these nuclear and astrophysical effects from dynamics in analyzing data. (Some recent papers emphasizing the important role of the velocity distribution, among other uncertainties, were refs. \cite{Lang:2010cd, Alves:2010pt}.)

Finally, the interaction that triggers signals in DM direct detection may leave imprints in hadron colliders like the Tevatron and LHC. The effective theory differs for these two types of experiments due to the different kinematic regimes that they probe. Still, for certain interactions, there may exist strong correlations between the direct detection and collider signals. Then it will be highly desirable to extract and compare information on DM dynamics from both classes of experiments.

\acknowledgments
We thank Nima Arkani-Hamed, Xiangdong Ji, Jay Wacker, and Neal Weiner for useful discussions. We also thank Marat Freytsis and Wai-Yee Keung for pointing out mistakes in Appendix C.3 in the previous version. M.R. thanks the PCTS for its support and the SLAC theory group for its hospitality while a portion of this work was completed. M.R. and L.-T.W. thank the Aspen Center for Physics for its hospitality while this work was in progress. L.-T.W. is supported by the National Science Foundation under grant PHY-0756966 and the Department of Energy (D.O.E.) under Outstanding Junior Investigator award DE-FG02-90ER40542.  

\appendix 

\section{Direct detection recoil rate}
\label{app: direct detection}

The basic quantity that  direct detection (except for bubble chamber type experiment which is mainly senstive to the total rate) measures is the differential event rate per unit recoiling energy
\beq 
 \frac{dR}{dE_R} =N_T \frac{\rho_\chi}{m_\chi}\int_{v_{min}}^{v_E} d^3v \, v f(v,v_E) \frac{d\sigma}{d E_R}. 
 \label{rate}
\eeq
$N_T$ is the number of scattering centers per unit detector mass. $v$ is the DM speed relative to the target nucleus and $\rho_\chi$ is the local halo density of DM particle near the Earth, which we take to be $\rho_\chi \approx 0.3$ GeV/cm$^3$. Then $v n_\chi = v \rho_\chi/m_\chi$ is the incident WIMP flux, and $f(v,v_E)$ is the WIMP velocity distribution in the galactic halo with velocity in the range $500\,{\rm km}/{\rm s}\, \le v_E \le 600\,$km/s. In the numerical studies, we take the distribution to be Maxwellian with $v_E = 500$ km/s, and average over modulation due to the solar system and Earth velocities \cite{Lewin:1995rx,Dehnen:1997cq} during the dates the experiment ran, when reported, or over the whole year when not reported. The minimum velocity to scatter with recoiling energy $E_R$ is 
\beq
v_{min} = \frac{\sqrt{2m_NE_R}}{2\mu_N} + \frac{\delta}{\sqrt{2m_NE_R}},
\eeq
where $m_N$ is the nucleus mass; $\mu_N$ is the DM--nucleus reduced mass and for the inelastic DM scenario, $\delta$, usually a mass splitting between DM components, characterizes the energy lost in the scattering to the dark sector. The total number of counts then follows from integrating over the energy bins, and multiplying by the effective exposure, e.g. in kg-days, for a given detector. Throughout the paper, we focus on elastic scattering and set $\delta=0$.

The form of the counting rate (\ref{rate}) isolates the physics of the WIMP-nucleus interaction in the
differential cross section $d\sigma/d E_R$. This differential cross section is computed from the NR matrix element as (averaging and summing over initial- and final-state spins):
\beq
\frac{d\sigma}{dE_R} = \frac{m_N}{2\pi v^2} \frac{1}{(2J + 1)(2 s_\chi + 1)} \sum_{\rm spins} |{\cal M}_{NR}|^2.
\eeq
We have denoted the spin of the nucleus as $J$. As mentioned in section \ref{sec:NReffpot}, for concreteness we will specialize to the case that $\chi$ is a Dirac fermion, so $s_\chi = 1/2$.

Combining Eq.~(\ref{eq: matrix element}), Eq.~(\ref{eq: Hamiltonian}), Eq.~(\ref{eq: Hamiltonian sd}), we have the recoil rate in terms of the parameters of NR effective theory as:
\beqs
 \frac{dR}{dE_R}& =&N_T \frac{\rho_\chi}{m_\chi}\int_{v_{min}}^{v_E} d^3v \, v f(v,v_E) \frac{d\sigma}{d E_R} \nonumber \\
\frac{d\sigma^{SI}}{dE_R}&=&\frac{A^2 F^2(E_R)m_N}{2\pi v^2}\left(\left|h_1 + \frac{l_1}{2 m_N E_R}\right|^2 + \frac{1}{4}\left|h_2 \sqrt{2 m_N E_R} + \frac{l_2}{\sqrt{2 m_N E_R}}\right|^2\right) \nonumber \\
\frac{d\sigma^{SD}}{dE_R}&=&\frac{J(J+1)m_N}{2\pi v^2}\frac{S(E_R)}{S(0)}\left(\frac{1}{4}\left|h_1^\prime + \frac{l_1^\prime}{2 m_N E_R}\right|^2 + \frac{1}{3}\left|h_2^\prime \sqrt{2 m_N E_R} + \frac{l_2^\prime}{\sqrt{2 m_N E_R}}\right|^2\right).\nonumber \\
\label{eqs: rate}
\eeqs
Note that the terms in $V_{\rm eff}$ do not all interfere with one another, due to their differing spin structure.

\section{Power counting}
\label{sec: powercounting}
In this Appendix, we do some simple power counting exercises for the two limits of mediators to show that the leading momentum suppressed operators could still be relevant to direct detection. As discussed in Sec.~\ref{sec: general}, there are at most four scales relevant to DM scattering off nucleus: DM mass $m_\chi$, nucleus mass $m_N$, mediator mass $m_\phi$ and some hidden sector scale $\Lambda^\prime$ unrelated to the first three. The NR expansion parameters fall into two classes: $v, q/m_\chi, q/m_N \sim 10^{-3}$ for DM mass around the weak scale, and $q/m_\phi, q/\Lambda^\prime$ which are unfixed. As argued in the context of momentum dependent DM, $q/m_\phi$ could be as large as 0.1 as in the GeV dark sector models. Notice that as we argued in Sec.~\ref{sec:nuclear}, there could not be large expansion parameters such as the {\em nucleon} velocity $v_n \sim 0.1$ present.

For elastic SI scattering, the smallest DM--nucleon cross section probed by the present direct detection experiments is around $\sigma_{p} \sim (10^{-43} -10^{-44})$ cm$^2$. Future upgraded experiments such as Xenon1T will probe cross sections down to $10^{-47}$ cm$^2$, close to $10^{-48}$ cm$^2$, the irreducible neutrino-induced nucleus recoil background.  For elastic SD scattering, the current bound on cross section is around the same order of magnitude of that of the weak interactions, $\sigma_p \sim (10^{-37} - 10^{-38})$ cm$^2$. (Despite the orders of magnitude separating these limits, it is not at all unnatural to expect both SI and SD signals; for instance, this is a very robust expectation if the DM is the MSSM LSP \cite{Cohen:2010gj}.) The bounds on the averaged DM--nucleus cross section are
\beqs
 \label{eq:bounds}
\langle\sigma_{N}^{SI} \rangle &\sim& A^{-2}\sigma_{N}^{SI} \sim {\mu_N^2\over \mu_n^2}
\sigma_{p}^{SI}  \nonumber \\
&=& \left(\frac{1 {\gev}}{\mu_n}\right)^2\left(\frac{\mu_N}{100 {\gev}}\right)^2 (10^{-39} - 10^{-40}) cm^2 \nonumber \\
&=&\left(\frac{1 {\gev}}{\mu_n}\right)^2\left(\frac{\mu_N}{100 {\gev}}\right)^2 \left({1 \over 200 {\gev}}\right)^2 (10^{-7} -10 ^{-8})~; \\
\langle\sigma_{N}^{SD} \rangle &\sim&{\mu_N^2\over \mu_n^2}\sigma_{p}^{SD}\nonumber  \\
 &=& \left(\frac{1 {\gev}}{\mu_n}\right)^2\left(\frac{\mu_N}{100 {\gev}}\right)^2 (10^{-33} - 10^{-34}) cm^2 \nonumber \\
 &=&\left(\frac{1 {\gev}}{\mu_n}\right)^2\left(\frac{\mu_N}{100 {\gev}}\right)^2 \left({1 \over 200 {\gev}}\right)^2(10^{-1}- 10^{-2})~.
\eeqs
where we defined the averaged DM--nucleus cross section to get rid of the multiplicity factor of atomic number $A$ or total charge $Z$. 

Now we do a crude power counting exercise to estimate to what order the operators could still contribute to direct detection: 
\begin{itemize}
\item{Heavy mediator: $m_\phi \sim  \calo(100 \,\gev)$. 
\beqs
i {\cal{M}}_{SI} &\sim& \tih_1 + \tih_2\calo\left(\frac{|\vec{p}|, |\vec{q}| }{\Lambda}\right), \nonumber \\
i {\cal{M}}_{SD} &\sim& \tih^{\prime}_{1},
\eeqs
where the dimensionless coefficient $\tih(\tih^\prime)=g_\chi f_{\rm{nucl}}$ represents products of DM and nucleon coupling to the mediator. From Eq.~(\ref{eq:bounds}), we see that to have experimentally accessible cross sections, 
\beq
\tih_1 \sim 10^{-4},\, \tih_2 \sim 0.1, \,\tih_1^\prime \sim 0.1.
\eeq}

\item{Light mediator: $m_\phi^2 \ll q^2$. 
\beqs
i {\cal{M}}_{SI} &\sim& \frac{\Lambda^2}{q^2}\left( \til_1+ \dots +\til_4\calo\left(\frac{|\vec{p}|, |\vec{q}| }{\Lambda}\right)^3\right), \nonumber \\
i {\cal{M}}_{SD} &\sim& \frac{\Lambda^2}{q^2}\left(\til^{\prime}_1 + \dots +\til_3^{\prime}\calo\left(\frac{|\vec{p}|, |\vec{q}| }{\Lambda}\right)^2\right),
\eeqs
where $\Lambda$ collectively denotes DM mass or nucleus mass around 100 GeV. Here the couplings $\til_1 \sim 10^{-10},\,\til_{i+1} \sim 10^{3i}\,\til_1, \,\til^\prime_0 \sim 10^{-7},\, \til_{i+1}^\prime \sim 10^{3i}\,\til_0^\prime$ are relevant for direct detection.}
\end{itemize}
In both cases we truncate the expansion when the coefficient has to be of order one to be relevant for direct detection. More accurate bounds on the coefficients have been presented in Sec.~\ref{sec: numerics}. 

\section{Matching to effective field theory operators}
In this appendix, we will discuss the general rules to write down NR effective operators and the matching between the effective field theory operators for DM with different spins and the NR effective theory. 
\subsection{Fermionic DM}
\label{sec: fermion}
\subsubsection{Complete set of NR operators}
\label{sec: complete set}
In the center of mass frame, the amplitude of fermionic DM scattering off a nucleus can be expressed in terms of scalar invariants formed out of four independent 3-vectors: the transferred momentum $\vec{q}$, the DM incoming velocity $\vec{v}$ relative to the target and the DM and nucleus spins $\vec{s}_\chi$ and $\vec{s}_N$. In fact, it is somewhat more natural to parametrize the operators not in terms of $\vec{v}$ but instead in terms of the sum of momenta before and after scattering, $P^\mu = (p+p')^\mu = (2p+q)^\mu$, which is related to the velocity by $\vec{P} = 2\mu_N \vec{v} + \vec{q}$. With two spins and two momenta, it is found in~\cite{Dobrescu:2006au} that 16 independent rotationally invariant operators can be constructed and they include all possible spin configurations. Any other scalar operator involving at least one of the two spins can be expressed as a linear combination of the 16 operators with spin-independent coefficients that may depend on the momenta through $|\vec{q}|^2, |\vec{v}|^2$. Below is the list of the complete set in the momentum space without mediator propagators, with superscripts denoting $P$ and $C$ parity. In particular, notice that there are six CP-violating operators: $\calo_2$, $\calo_6$, $\calo_9$, $\calo_{13}$, $\calo_{14}$, and $\calo_{15}$.

\begin{itemize}
\item{SI operators
\beqs
\calo_1^{(++)}&=&1 \nonumber \\
\calo_2^{(-+)}&=&i\vec{s}_\chi \cdot \vec{q} \nonumber \\
\calo_3^{(--)}&=&\vec{s}_\chi \cdot \vec{P} \nonumber \\
\calo_4^{(++)}&=&i\vec{s}_\chi \cdot (\vec{P} \times \vec{q}) 
\label{eq: si nr op}
\eeqs
}

\item{SD operators
\beq
\begin{tabular}{lc|cl}
$\calo_5^{(++)}=\vec{s}_\chi \cdot \vec{s}_N$ &&& $\calo_{11}^{(++)}=(\vec{s}_N \cdot \vec{q})(\vec{s}_\chi \cdot \vec{q})$\\
$\calo_6^{(-+)}=i\vec{s}_N \cdot \vec{q}$ &&& $\calo_{12}^{(++)}=(\vec{s}_N \cdot \vec{P})(\vec{s}_\chi \cdot \vec{P})$\\
$\calo_7^{(--)}=\vec{s}_N \cdot \vec{P} $&&&$\calo_{13}^{(+-)}=i \left[(\vec{s}_N \cdot \vec{q})(\vec{s}_\chi \cdot \vec{P})+ (\vec{s}_N \cdot \vec{P})(\vec{s}_\chi \cdot \vec{q})\right] $\\
$\calo_8^{(--)}=i (\vec{s}_\chi \times \vec{s}_N)\cdot \vec{q} $ &&& $\calo_{14}^{(+-)}=i \left[(\vec{s}_N \cdot \vec{q})(\vec{s}_\chi \cdot \vec{P}) - (\vec{s}_N \cdot \vec{P})(\vec{s}_\chi \cdot \vec{q})\right] $\\
$\calo_9^{(-+)}=(\vec{s}_\chi \times \vec{s}_N)\cdot \vec{P}$ &&& $\calo_{15}^{(-+)}=\left[ \vec{s}_N \cdot (\vec{P}\times\vec{q}) \right] (\vec{s}_\chi \cdot \vec{q})+\left[ \vec{s}_\chi \cdot (\vec{P}\times\vec{q}) \right] (\vec{s}_N \cdot \vec{q})$ \\
$\calo_{10}^{(++)}=i\vec{s}_N \cdot (\vec{P}\times\vec{q})$ &&& $\calo_{16}^{(--)}=i\left[ \vec{s}_N \cdot (\vec{P}\times\vec{q}) \right] (\vec{s}_\chi \cdot \vec{P})+\left[ \vec{s}_\chi \cdot (\vec{P}\times\vec{q}) \right] (\vec{s}_N \cdot \vec{P})$
\end{tabular}
 \label{eq: sd nr op}
\eeq}
\end{itemize}

Among the 16 operators, there are four nucleus SI operators ($\calo_1 - \calo_4$) with only $\calo_4$ suppressed by two powers of the expansion parameter $|\vec{v}|, |\vec{q}|$. In coordinate space,  $\calo_1 -\calo_3$ lead to 3 types of contact interaction for heavy mediator and 3 long-range interactions for light mediator. The 4 cases arising from the two operators $\calo_1, \calo_2$ (with at most a single $|\vec{q}|$ suppression) are included in our parametrization in Eq.~(\ref{eq: Hamiltonian}). 

The other 13 operators $\calo_5 - \calo_{16}$  are all SD. Up to leading order in $\vec{v}, \vec{q}$, one has $\calo_5 - \calo_9$. The amplitudes of $\calo_8, \calo_9$ are almost the same as those of $\calo_6, \calo_7$ except for different dependence on the DM spin. Thus, after averaging over initial spins and summing over final spins, $\calo_8 (\calo_9)$ give differential cross sections identical to those from $\calo_6 (\calo_7)$ up to a numerical factor. Thus we do not include them in Eq.~(\ref{eq: Hamiltonian sd}). 

The physical interpretations of the operators are more transparent in the potential form. For instance, if the mediator is light, $\calo_2, \calo_6$ are the dipole--monopole potentials while $\calo_{11}$ is the dipole--dipole potential.

\subsubsection{Field theory operators}
Now we turn to field theory operators and write down the most general four fermion interaction mediated by some scalar or gauge boson in the form:
\beq
G(q^2,m_{\phi}) J_{\chi}  J_{q} = G(q^2,m_\phi)\bar{\chi}\Gamma_\chi\chi\bar{q}\Gamma_qq,
\eeq
where $J_\chi$ and $J_q$ are the DM fermion and quark bilinears (we hope the reader will not be confused by our use of $q$ both for quarks and the momentum transfer). The propagator $G(q^2, m_\chi)$ is approximated as $1/m_{\phi}^2$ in the heavy mediator case and $1/q^2$ in the light mediator case. The DM fermion could be a single Majorana, Dirac or two Majoranas. $\Gamma_{\chi(q)}$ is a 4 by 4 matrix, which we choose to be $I, \gamma^5, \gamma^\mu,\gamma^\mu\gamma^5, \sigma^{\mu\nu}, \sigma^{\mu\nu}\gamma^5$ for non-derivative interactions. Correspondingly, we have the following DM and quark non-derivative interactions
\beqs
\rm{(pseudo)scalar \times (pseudo)scalar}&:&  \bar{\chi}(\gamma^5) \chi \bar{q}(\gamma_5) q \nonumber \\
\rm{(pseudo)vector \times (pseudo)vector}&:&  \bar{\chi}\gamma^\mu(\gamma^5) \chi \bar{q}\gamma_\mu(\gamma^5) q \nonumber \\
\rm{(pseudo)tensor \times (pseudo)tensor}&:& \bar{\chi} \sigma^{\mu\nu}(\gamma^5)\chi \bar{q}\sigma_{\mu\nu}(\gamma^5)q \nonumber. 
\eeqs
Notice that not all of them are independent. For instance, as $\sigma^{\mu\nu}\gamma^5 = \frac{i}{2} \epsilon^{\mu\nu\rho\sigma} \sigma_{\rho\sigma}$, the pseudo-tensor $\times$ pseudo-tensor interaction is equivalent to the tensor $\times$ tensor interaction and only one pseudo-tensor $\times$ tensor interaction is independent.  If the DM is a single Majorana fermion, operators such as $\bar{\chi}\gamma^\mu \chi$ and $\bar{\chi}\sigma^{\mu\nu}(\gamma^5) \chi$ do not exist. But in the NR limit, one cannot tell the difference between single Majorana and Dirac fermions as they lead to the same set of NR operators. Below we summarize in Table.~\ref{table:fermion DM} the correspondence between field theory operators and NR operators. We also list both contact and long-range potentials for every operator.

\begin{table}
\begin{center}
\begin{tabular}{|c|c|c|c|c|}
\hline 
 &Effective operator & leading NR operators & \multicolumn{2}{|c|}{ leading NR operators} \\
 && in momentum space &  \multicolumn{2}{|c|}{in position space} \\
\hline
\hline
& $\bar{\chi} \chi \bar{q}q$ & 1 & $\delta^3(\vec{r})$ & $\frac{1}{r}$ \\
&$i\bar{\chi} \gamma^5 \chi \bar{q}q$& $i\vec{s}_\chi \cdot \vec{q}$ & $-\vec{s}_{\chi}\cdot \vec{\nabla}\delta^3(\vec{r})$ &$\frac{\vec{s}_{\chi}\cdot \vec{r}}{ r^3}$ \\
SI&$\bar{\chi}\gamma^\mu \chi \bar{q}\gamma_\mu q$ & 1& $\delta^3(\vec{r})$&$\frac{1}{r}$\\
& $\bar{\chi} \gamma^5 \gamma^\mu\chi \bar{q}\gamma_{\mu}q$ & $\vec{s}_\chi^\bot\cdot \vec{v}$ &$(\vec{s}_{\chi}\cdot \vec{v}+ \frac{i}{2\mu_N}\vec{s}_\chi\cdot \vec{\nabla})\delta^3(\vec{r})$ &$\frac{\vec{s}_{\chi}\cdot \vec{v}}{r} - \frac{i\vec{s}_{\chi}\cdot \vec{r}}{2\mu_Nr^3}$\\
&$i\bar{\chi} \sigma^{\mu\nu}\gamma^5\chi \bar{q}\sigma_{\mu\nu}q$ & $i\vec{s}_\chi \cdot \vec{q}$ &  $-\vec{s}_{\chi}\cdot \vec{\nabla}\delta^3(\vec{r})$&$\frac{\vec{s}_{\chi}\cdot \vec{r}}{ r^3}$ \\
\hline
\hline
& $i\bar{\chi}\chi\bar{q}\gamma_5q$  &$i\vec{s}_N \cdot \vec{q}$& $-\vec{s}_N\cdot \vec{\nabla}\delta^3(\vec{r})$ &$\frac{\vec{s}_N\cdot \vec{r}}{ r^3}$ \\
  &$\bar{\chi} \gamma^5 \chi \bar{q}\gamma^5 q$ & $(\vec{s}_\chi \cdot \vec{q})(\vec{s}_N \cdot \vec{q})$& $(\vec{s}_\chi \cdot \vec{\nabla})(\vec{s}_N \cdot \vec{\nabla})\delta^3(\vec{r})$ &$\frac{3(\vec{s}_{\chi}\cdot \vec{r})(\vec{s}_N\cdot \vec{r})}{r^5} - \frac{\vec{s}_{\chi}\cdot \vec{s}_N}{r^3}$\\
  &$\bar{\chi} \gamma^5 \gamma^\mu\chi \bar{q}\gamma_{\mu}q$&$i(\vec{s}_\chi \times \vec{s}_N) \cdot \vec{q}$&$-(\vec{s}_{\chi}\times\vec{s}_N)\cdot\vec{\nabla}\delta^3(\vec{r})$ &$ { \vec{r}\cdot(\vec{s}_\chi \times \vec{s}_N)\over 4\pi r^3 }$ \\
 SD&$\bar{\chi}\gamma^\mu\chi \bar{q}\gamma_{\mu} \gamma^5q$  &  $\vec{s}_N^\bot\cdot \vec{v}$ &$(\vec{s}_N\cdot \vec{v}+ \frac{i}{2\mu_N}\vec{s}_N\cdot \vec{\nabla})\delta^3(\vec{r})$ & $\frac{\vec{s}_N\cdot \vec{v}}{r} - \frac{i\vec{s}_N\cdot \vec{r}}{2\mu_Nr^3}$\\
 &&$i(\vec{s}_\chi \times \vec{s}_N) \cdot \vec{q}$&$-(\vec{s}_{\chi}\times\vec{s}_N)\cdot\vec{\nabla}\delta^3(\vec{r})$ &$ { \vec{r}\cdot(\vec{s}_\chi \times \vec{s}_N)\over 4\pi r^3 }$ \\
  &$\bar{\chi}\gamma^5\gamma^\mu\chi \bar{q}\gamma_{\mu} \gamma^5q$ &$\vec{s}_{\chi}\cdot \vec{s}_N$ &$\vec{s}_{\chi}\cdot \vec{s}_N\delta^3(\vec{r})$&$\frac{\vec{s}_{\chi}\cdot \vec{s}_N}{ r}$\\
 &$\bar{\chi} \sigma^{\mu\nu}\chi \bar{q}\sigma_{\mu\nu}q$ &$\vec{s}_{\chi}\cdot \vec{s}_N$&$\vec{s}_{\chi}\cdot \vec{s}_N\delta^3(\vec{r})$&$\frac{\vec{s}_{\chi}\cdot \vec{s}_N}{r}$\\
\hline
\end{tabular}
\caption{Effective operators of fermion DM and their corresponding NR operators}
\label{table:fermion DM}
\end{center}
\end{table}

A few comments are in order:
\begin{itemize}
\item{We exhausted all possible four fermion non-derivative interactions, but the NR leading operators in the momentum space only cover 8 of the complete set of the 16 operators listed in Eq.~(\ref{eq: si nr op}, \ref{eq: sd nr op}). In order to get the other operators, one needs to combine and manipulate the effective operators listed above so that the NR leading terms cancel out. For instance, to get the SI operator ${\cal{O}}_4 = i\vec{s}_\chi \cdot (\vec{P} \times \vec{q})$, tuning is necessary to cancel a leading operator. (Note that, e.g., an expression that Fierzes to $\bar{\chi}\chi\bar{q}q-\bar{\chi}\gamma^\mu\chi \bar{q}\gamma_\mu q$ does not lead to an automatic cancellation, because the first term depends on $A$ times nuclear matrix elements whereas the second depends on $Z$ times different nuclear matrix elements.).}

\item{As noted above, the velocity appears through the combination $\vec{P} = 2\mu_N \vec{v} + \vec{q}$, with $\mu_N$ the DM--nucleus reduced mass. For instance, consider a single Majorana fermion coupling to the $Z$ boson. After integrating out the $Z$, the leading unsuppressed operator is SD, $\bar{\chi}\gamma^5\gamma^\mu \chi \bar{q}\gamma^5\gamma^\mu q$, giving a potential $\vec{s}_{\chi}\cdot \vec{s}_N\delta^3(\vec{r})$. The leading SI operator is $\bar{\chi}\gamma^5\gamma^\mu \chi \bar{q}\gamma_\mu q$, giving $V \sim (\vec{s}_{\chi}\cdot \vec{v}+ \frac{i}{2\mu_N}\vec{s}_\chi\cdot \vec{\nabla})\delta^3(\vec{r})\sim \vec{s}_\chi^\bot \cdot \vec{v}\delta^3(\vec{r})$ where $\vec{s}^\bot$ is the spin transverse to the momentum transfer. This is known as DM ``anapole" moment scattering off the nuclear electric current. The SI scattering cross section is suppressed by $(q/m_Z)^2 \sim 10^{-6}$. But notice that the current SD bound from direct detection is about $10^{-38}$ cm$^2$, six orders of magnitude above the SI bound $10^{-44}$ cm$^2$. Thus with all couplings of order one, both operators are relevant for direct detection. This is an example where one cannot naively discard a momentum-suppressed SI operator if the leading operator is SD.}

\item{Derivative effective field operators could also give NR operators in Eq.~(\ref{eq: Hamiltonian}, \ref{eq: Hamiltonian sd}). These operators arise naturally from dark moment models. For instance, the DM electric dipole moment scattering off the nucleus charge is encoded in the matrix element 
\beq
\frac{\bar{\chi} \sigma^{\mu\nu}D_\mu \gamma_5\chi \bar{q}\gamma_\nu q}{\vec{q}^2}
\eeq
In the NR limit, it yields the (DM) dipole coupling $\vec{s}_\chi \cdot \vec{r}/ r^3$.
The DM magnetic moment scatters off both the nucleus charge and spin, leading to NR operators $\vec{s}_\chi \cdot \vec{s}_N$ and $\vec{s}_\chi \cdot (\vec{v} \times \vec{q})$ as discussed in the end of Sec.~\ref{sec: nucleon currents}.
}

\end{itemize}

\subsubsection{Examples of matching}
\label{sec:example}
{\bf Pseudoscalar Exchange} \\
We consider a light pseudoscalar coupled to the DM and visible sector as
\beq
i \alpha \phi \bar{\chi} \gamma^5 \chi + \beta \phi \bar{q} q +h.c,
\eeq
where we assume that $\beta$ is flavor diagonal. The coupling to quarks could be translated to that to the proton as
\beq
f_p=\sum_{q=u,d,s} \frac{m_p}{m_q}\beta_q f_{T_q}^{(p)} + \frac{2}{27}\left(1-\sum_{q=u,d,s}f_{T_q}^{(p)}\right)\sum_{q=c,b,t}\beta_q \frac{m_p}{m_q}~.
\eeq
Then the coupling to the nucleus is $f_N = Z f_p + (A-Z) f_n$ with form factor set aside for the moment.

In the NR limit, the matrix element is 
\beq
{\cal{M}}^{SI}=i\alpha f_N 4 m_N \vec{q} \cdot \vec{s}_\chi.
\eeq
After taking into account different normalizations, the NR matrix element is related to the relativistic matrix element as 
\beq
{\cal{M}}_{NR} = \frac{{\cal{M}}}{4m_\chi m_N}.
\eeq
The corresponding NR effective potential is the Fourier transform of the NR matrix element to the leading order of Born approximation
\beqs
V^{SI}& =& l_2 \frac{\vec{s}_\chi\cdot \vec{r}}{4\pi r^3}, \nonumber \\
{\rm{where}} \quad l_2 &=& \frac{\alpha}{m_\chi} \left(\frac{Z}{A} f_p+ \frac{Z-A}{A} f_n\right).
\eeqs
The differential recoil rate is then
\beq
\frac{d\sigma^{SI}}{dE_R}=\frac{A^2 F^2(E_R)m_N}{8\pi v^2} \left|\frac{l_2}{\sqrt{2 m_N E_R}}\right|^2.
\eeq
\\
\noindent
{\bf DM anapole moment scattering}\\
Now we will move to a more elaborate example of the NR effective theory, which goes beyond the small set of operators that we considered in the main text. The model we consider has a Majorana fermion as DM. It interacts with the nucleus through the $Z$ boson. This leads to the anapole moment scattering which has been considered recently, for example, in \cite{Fitzpatrick:2010br}. After integrating out the $Z$, the effective theory operator for the DM--quark interaction is 
\beq
\alpha\bar{\chi}\gamma^\mu\gamma_5\chi\bar{q}\gamma_\mu q+\beta \bar{\chi}\gamma^\mu\gamma_5\chi\bar{q}\gamma_\mu\gamma_5 q,
\eeq
where the coefficients $\alpha, \beta$ are 
\beqs
\alpha_u &=& \frac{g_\chi \sqrt[4]{2}G_F^{1/2}}{m_Z}\left(\frac{1}{2}-\frac{4}{3}\sin^2\theta_W\right) \nonumber \\
\alpha_d &=& \frac{g_\chi \sqrt[4]{2}G_F^{1/2}}{m_Z}\left(-\frac{1}{2}+\frac{2}{3}\sin^2\theta_W\right) \nonumber \\
\beta_u &=&- \frac{g_\chi \sqrt[4]{2}G_F^{1/2}}{2m_Z}\nonumber \\
\beta_d &=& \frac{g_\chi \sqrt[4]{2}G_F^{1/2}}{2m_Z}
\eeqs
where the subscripts of $\alpha, \beta$ denote the couplings of up or down type quarks. $g_\chi$, the DM--$Z$ coupling constant, is a free parameter. (As mentioned in Sec.~\ref{sec: models}, it can arise from a higher-dimension operator such as $\bar{\chi}\gamma^\mu \chi h^\dagger D_\mu h$, and so is not directly related to the SM gauge couplings $g$ and $g'$.)
Then we use the recipe outlined in Sec.~\ref{sec:nuclear} to map the DM--quark operator to DM--nucleus operator. At the level of {\em nucleons}, we have couplings:
\beq
\bar{\chi}\gamma^\mu\gamma_5\chi\left({\cal J}_\mu + {\cal A}_\mu\right) \equiv \bar{\chi}\gamma^\mu\gamma_5\chi\left(f_p \bar{p}\gamma_\mu p + f_n \bar{n}\gamma_\mu n  + a_p \bar{p}\gamma_\mu \gamma_5 p + a_n \bar{n}\gamma_\mu \gamma_5 n \right)~, \nonumber \\
\eeq
Where we have denoted the vector and axial currents on the right-hand side by ${\cal J}$ and ${\cal A}$ and defined:
\beqs
f_p &=& 2\alpha_u + \alpha_d, \nonumber \\
f_n &=& \alpha_u+ 2 \alpha_d, \nonumber \\
a_p &=& \sum_{q\,=\,u, d, s} \Delta_q^{(p)}\beta_q, \nonumber  \\
a_n &=& \sum_{q\,=\,u, d, s} \Delta_q^{(n)}\beta_q,
\eeqs
with $\Delta_q^{(p)}$ the fraction of spin of the proton carried by quark flavor $q$, as reviewed in Sec.~\ref{sec: match to nucleon}. 

Now, we evaluate matrix elements in the nucleus:
\beqs
\bar{\chi}\gamma^\mu \gamma_5 \chi \left<N\middle|{\cal J}_\mu\middle|N\right> & = &  \frac{Z f_p  + (A-Z) f_n}{2 m_N A} \bar{\chi}\gamma^\mu \gamma_5 \chi \bar{N}\left(A F(q)\left(p+p'\right)_\mu + 2i\Sigma_{\mu\nu}q^\nu \tilde{F}(q)\right) N\nonumber \\
\bar{\chi}\gamma^\mu \gamma_5 \chi \left<N\middle|{\cal A}_\mu\middle|N\right> & = & \sqrt{\frac{S(q)}{S(0)}} \left(a_p\frac{\langle S_p \rangle} {J} + a_n \frac{\langle S_n \rangle} {J}\right) \, \bar{\chi}\gamma^\mu\gamma_5 \chi \bar{N} (2S_\mu) N.
\eeqs
Here we have defined $A F(q)$ and $\tilde{F}(q)$ to be the charge and magnetic dipole moment form factors for the baryon number current (we have assumed that protons and neutrons are distributed uniformly throughout the nucleus). $S_\mu$ denotes the nucleus spin. As in Sec.~\ref{sec: nucleon currents}, $\bar{N}$ and $N$ denote appropriate spin-$J$ wave functions and $\Sigma_{\mu\nu}$ a spin-$J$ Lorentz representation. The two terms arising from ${\cal J}$ do not mix with each other or with the term arising from ${\cal A}$, so will discuss the three separately.

For convenience, we will define constants:
\beqs
c_{\cal J} & = & \frac{Z f_p + (A-Z) f_n}{A}, \nonumber \\
c_{\cal A} & = &  a_p\frac{\langle S_p \rangle} {J} + a_n \frac{\langle S_n \rangle} {J}.
\eeqs 
Now we will calculate the NR matrix elements. First, from ${\cal J}$ we extract the piece proportional to $F(q)$. A straightforward calculation shows that it gives a spin-independent matrix element. Here we factor out $A F(q)$, according to our convention for SI NR matrix elements as set out in Sec.~\ref{sec: general}. (Recall that we have assumed $\chi$ to be Majorana; certain factors of 2 would differ if it were Dirac.)
\beqs
{\cal M}^{SI} & = &16 m_\chi m_N c_{\cal J} \left(\vec{v} + \frac{\vec{q}}{2\mu_N}\right) \cdot \vec{s}_\chi \nonumber \\
& = & 16 m_\chi m_N c_{\cal J} (\vec{s}_\chi-\hat{q}\cdot\vec{s}_\chi\,\hat{q}) \cdot \vec{v}  \equiv 16 m_\chi m_N c_{\cal J} \vec{s}_\chi^\bot \cdot \vec{v}.
\eeqs
Next, we consider the spin-dependent piece arising from the magnetic dipole moment term in the matrix element of ${\cal J}$. Because it is spin-dependent, we follow our convention of factoring out  $\sqrt{\frac{S(q)}{S(0)}}$, even though the form factor in this case, $\tilde{F}(q)$, is different. This leads to some awkwardness in the definition of our coupling:
\beqs
{\cal M}^{SD}_{\cal J} = 16ic_{\cal J}\tilde{F}(q)\sqrt{\frac{S(0)}{S(q)}} m_\chi \vec{s}_\chi \cdot \left(\vec{q} \times \vec{s}_N\right).
\eeqs
Notice that this matrix element corresponds to the operator ${\cal O}_8$ in the general classification of App.~\ref{sec: complete set}. Finally, we have the spin-dependent matrix element arising from the axial current ${\cal A}$:
\beqs
{\cal M}^{SD}_{\cal J} = 32 c_{\cal A} m_\chi m_N \vec{s}_\chi \cdot \vec{s}_N.
\eeqs
Then, dividing by $4 m_\chi m_N$ to convert to the nonrelativistic normalization, we obtain the corresponding NR effective potentials:
\beqs
V^{SI} & = & \left(h_1^{(v)} \vec{v} \cdot \vec{s}_\chi + i h_2 \vec{\nabla} \cdot \vec{s}_\chi\right) \delta^3(\vec{r}), \nonumber \\
V^{SD} &=& h_1^\prime \,\vec{s}_N \cdot \vec{s}_\chi \delta^3(\vec{r})~ + h_8 \, \left(\vec{s}_\chi \times \vec{s}_N\right) \cdot \vec{\nabla}\delta^3(\vec{r}), 
\eeqs
where
\beqs
h_1^{(v)} = 2 \mu_N h_2 & = & 4 \left(\frac{Z f_p + (A-Z) f_n}{A} \right) \nonumber \\
h_1^\prime &=& 8 \left(\frac{\langle S_p \rangle} {J} + a_n \frac{\langle S_n \rangle} {J}\right) \nonumber \\
h_8 & = & -4  \left(\frac{Z f_p + (A-Z) f_n}{A} \right) \tilde{F}(q) \sqrt{\frac{S(0)}{S(q)}}.
\eeqs

The differential recoil rate is then
\beqs
\frac{d\sigma^{SI}}{dE_R}&=&\frac{A^2 F^2(E_R)m_N}{2\pi v^2}\left(\left|\frac{1}{2}h^{(v)}_1 v+ \frac{1}{2}h_2 \sqrt{2 m_N E_R} \right|^2\right)~, \nonumber \\
\frac{d\sigma^{SD}}{dE_R}&=&\frac{J(J+1)m_N}{2\pi v^2}\frac{S(E_R)}{S(0)}\left(\frac{1}{4}\left|h_1^\prime \right|^2 + \frac{1}{12} \left|h_8 \sqrt{2 m_N E_R}\right|^2\right).
\eeqs
Notice that the SI rate is slightly different from Eq. (\ref{eqs: rate}) as it includes the interference term between the velocity-suppressed and momentum-suppressed operators. Furthermore, the SD rate includes a term very similar to the familiar $h_2^\prime$ contribution, but arising instead from the operator ${\cal O}_8 = i (\vec{s}_\chi \times \vec{s}_N)\cdot \vec{q}$. Thus, this example goes beyond the formalism of App.~\ref{app: direct detection}. We have avoided such subtleties in most of the text because the limited set of coefficients $h_1$, $h_1^\prime$, $l_1$, $l_1^\prime$ suffice to produce most of the interesting variations in recoil spectrum shape. Our new $h_1^{(v)}$ term gives a shape similar to that arising from $h_1$, while $h_8$ gives a shape similar to that arising from $h_2^\prime$. The presence of these new coefficients in an example as natural as $Z$ exchange, however, shows that the inverse problem of matching a field theory to a measured shape is more subtle, and necessarily involves a number of degeneracies due to the higher-dimensional space of possible potentials that yield similar shapes.

\subsection{Scalar DM}
\label{sec: scalar}
The analysis for the scalar DM is greatly simplified due to the trivial fact that scalar has no spin. In the momentum space, only three NR operators could be constructed from $\vec{q}, \vec{v}, \vec{s}_N$ to the order of $|\vec{q}|, |\vec{v}|$ :
\beqs
\calo_{SI} &=& 1 \nonumber \\
\calo_{SD} &=&  \vec{s}_N\cdot \vec{q} + \vec{s}_N\cdot \vec{P}  \nonumber 
\eeqs
As we only consider static potential, only the first two operators remain which, in the position space, are translated to
\beqs
V_{SI} &=& \delta^3(\vec{r}),  \quad \frac{1}{4\pi r} \nonumber \\
V_{SD} &=&\vec{s}_N\cdot \vec{\nabla}\delta^3(\vec{r}),\quad \frac{ i\vec{s}_N\cdot \vec{r}}{4\pi r^3} 
\eeqs

In the field theory, one could write down scalar operators that couple to the (axial) scalar and  (axial) vector quark current:
\beq
\phi^2 \bar{q} (\gamma^5)q, \quad \phi^\dagger \partial_\mu \phi \bar{q} \gamma_\mu(\gamma^5)q.
\eeq
They are related to the NR operators in Table~\ref{table: scalar DM}. 
\begin{table}[h]
\begin{center}
\begin{tabular}{|c|c|c|c|c|}
\hline 
& Effective operator & leading NR operator & \multicolumn{2}{|c|}{ leading NR operators} \\
&& in momentum space& \multicolumn{2}{|c|}{in position space} \\
\hline
\hline
SI & $\phi^2 \bar{q} q, \quad \phi^\dagger \partial_\mu \phi \bar{q} \gamma_\mu q$ & 1 & $\delta^3(\vec{r})$&  $\frac{1}{4\pi r}$ \\
\hline
SD&$\phi^2 \bar{q} \gamma^5 q$ &  $\vec{s}_N\cdot \vec{q}$& $\vec{s}_N\cdot \vec{\nabla}\delta^3(\vec{r})$ &$\frac{ i\vec{s}_N\cdot \vec{r}}{4\pi r^3} $ \\
& $\phi^\dagger \partial_\mu \phi \bar{q} \gamma_\mu\gamma^5q$ &$\vec{s}_N^\bot\cdot \vec{v} $&  $(\vec{s}_N\cdot \vec{v}+ \frac{i}{2\mu_N}\vec{s}_N\cdot \vec{\nabla})\delta^3(\vec{r})$ & $\frac{\vec{s}_N\cdot \vec{v}}{r} - \frac{i\vec{s}_N\cdot \vec{r}}{2\mu_Nr^3}$\\
\hline

\end{tabular}
\caption{Effective operators of scalar DM and their corresponding NR operators.}
\label{table: scalar DM}
\end{center}
\end{table}

\subsection{Vector DM}
\label{sec: vector}
The spin-1 DM NR operator analysis is analogous to the fermionic DM. One modification is the representation of the spin operator. In the NR limit, vector DM spin operator is 
\beqs
\vec{s}_B &=& \epsilon^\dagger \cdot \vec{R} \cdot \epsilon \nonumber \\
{\rm{or}} \quad s_B^k &=& \epsilon^{ijk}\epsilon_i^*\epsilon_j
\eeqs
where $\epsilon^i$ is the 3 component polarization vector and $\vec{R}$ is 3-dimensional spin-1 rotation represenation with $(R^k)_{ij} = \epsilon^{ijk}$. In addition, besides the vector representation, high-dimensional tensor representation of the rotation group could be relevant  (one do not need to consider this for the spin $1/2$ case as the Pauli matrices form a complete basis for the 2 by 2 matrice). For instance, as we will show below, the symmetric representation with two indices could be relevant
\beq
{\cal{S}}^{ij}= \epsilon^\dagger \cdot R^{(i}R^{j)}\cdot \epsilon = \epsilon^{\dagger (i}\epsilon^{j)}.
\eeq

One prototype example of vector DM in the field theory is the the first KK mode of the photon in the universal extra dimension model \cite{Cheng:2002ej}. Again, one can write down operators involving the heavy vector DM $B^\mu$ that couple to the scalar and vector quark current. First if $B^\mu$ is self-conjugate, 
\beqs
&B^{\mu} B_\mu \bar{q} (\gamma^5)q,&  \nonumber \\
&B^{\mu} \partial_\mu B_\nu \bar{q}\gamma^\nu (\gamma^5)q&, \epsilon_{\mu\nu\rho\sigma}B^{\nu}\partial^\rho B^\sigma \bar{q} \gamma^\mu (\gamma^5)q. 
\eeqs
Additional operators exist if the gauge field is complex:
\beq
B^{\dagger\mu} \partial_\nu B_\mu \bar{q} \gamma^\nu (\gamma^5)q, 
\eeq
These operators are not gauge invariant and are generated from UV theory after integrating out heavy degrees of freedom. Thus we only consider contact interactions. In the NR limit, they lead to operators in table~\ref{table: vector DM}. 
\begin{table}[h]
\begin{center}
\begin{tabular}{|c|c|c|c|}
\hline 
& Effective operator & leading NR operator & leading NR operator \\
&& in momentum space&in position space \\
\hline
\hline
SI & $B^{\mu} B_\mu \bar{q} q, B^{\dagger\mu} \partial_\nu B_\mu \bar{q} \gamma^\nu q$  & 1 & $\delta^3(\vec{r})$ \\
&$ \epsilon_{\mu\nu\rho\sigma}B^{\nu}\partial^\rho B^\sigma \bar{q} \gamma^\mu q$ & $\vec{s}_B^\bot\cdot \vec{v} $ &  $(\vec{s}_B\cdot \vec{v}+ \frac{i}{2\mu_N}\vec{s}_B\cdot \vec{\nabla})\delta^3(\vec{r})$ \\
&$B^{\mu} \partial_\mu B_\nu \bar{q}\gamma^\nu q$ & $i\vec{P}\cdot {\cal{S}} \cdot \vec{q}$ & $-(\vec{v}+\frac{i}{2\mu_N}\vec{\nabla})\cdot{\cal{S}}\cdot\vec{\nabla}\delta^3(\vec{r})$\\
\hline
SD&$ B^{\mu} B_\mu \bar{q} \gamma^5q$ &$i\vec{q}\cdot \vec{s}_N$ &$-\vec{s}_N\cdot\vec{\nabla}\delta^3(\vec{r})$ \\
&$B^{\dagger\mu} \partial_\nu B_\mu \bar{q} \gamma^\nu \gamma^5q$&$\vec{s}_N^\bot \cdot \vec{v}$ &$(\vec{s}_N\cdot \vec{v}+ \frac{i}{2\mu_N}\vec{s}_N\cdot \vec{\nabla})\delta^3(\vec{r})$  \\
&$B^{\mu} \partial_\mu B_\nu \bar{q}\gamma^\nu \gamma^5q$& $i\vec{s}_N\cdot {\cal{S}} \cdot \vec{q}$ & $-\vec{s}_N\cdot {\cal{S}}\cdot\vec{\nabla}\delta^3(\vec{r})$\\
&$ \epsilon_{\mu\nu\rho\sigma}B^{\nu}\partial^\rho B^\sigma \bar{q} \gamma^\mu \gamma^5q$ &$\vec{s}_B\cdot \vec{s}_N$ &$\vec{s}_B\cdot \vec{s}_N\delta^3(\vec{r})$ \\
&$ \epsilon_{\mu\nu\rho\sigma}B^{\nu}\partial^\rho B^\sigma \bar{q} \gamma^\mu q$&$ i (\vec{s}_B \times \vec{s}_N) \cdot \vec{q}$ & $- (\vec{s}_B \times \vec{s}_N) \cdot \delta^3(\vec{r})$ \\
\hline

\end{tabular}
\caption{Effective operators of vector DM and their corresponding NR operators.}
\label{table: vector DM}
\end{center}
\end{table}

We will not explicitly consider higher-spin DM, although the story generalizes straightforwardly. A discussion of the spin-$3/2$ case can be found in \cite{Barger:2008qd}.

\end{document}